\documentclass[english,aps,preprint,nofootinbib]{revtex4}
\usepackage[T1]{fontenc}
\usepackage[latin1]{inputenc}
\usepackage{longtable}
\usepackage{amsmath}
\usepackage{graphicx}
\usepackage{amssymb}

\makeatletter

\providecommand{\tabularnewline}{\\}

\usepackage{bm}
\usepackage{slashed}
\usepackage{braket}
\usepackage{nicefrac}

\newcommand{\ssst}[1]{\scriptscriptstyle{#1}}
\newcommand{\wt}[1]{\widetilde{#1}}

\newcommand{\mN}{m_{\ssst{N}}}

\usepackage{multirow}
\usepackage{dcolumn}
\newcolumntype{d}[1]{D{.}{.}{#1}}

\usepackage{babel}
\makeatother
\begin{document}

\preprint{LA-UR-06-6771}

\title{Parity-Violating Observables of Two-Nucleon Systems in Effective
Field Theory}

\author{C.-P. Liu}

\email{cpliu@lanl.gov}

\affiliation{T-16, Theoretical Division, Los Alamos National Laboratory, Los Alamos,
NM 87545, USA}

\affiliation{Theory Group, Kernfysisch Versneller Instituut, University of Groningen,
Zernikelaan 25, 9747 AA Groningen, The Netherlands}

\begin{abstract}
A newly-proposed parity-violating nucleon-nucleon interaction based
on effective field theory is studied in this work. It is found that
at low energy, i.e., $E_{\mathrm{lab}}\lesssim90\,\mbox{MeV}$ for
$^{1}S_{0}$--$^{3}P_{0}$ transitions and $E_{\mathrm{lab}}\lesssim40\,\mbox{MeV}$
for $^{3}S_{1}$--$^{1}P_{1}$ and $^{3}S_{1}$--$^{3}P_{1}$ transitions,
the parity-violating observables can be completely specified by a
minimal set of six parameters. It contains five low-energy constants,
which are equivalent to the Danilov parameters, and an additional
parameter that characterizes the long-range one-pion exchange and
is proportional to the parity-violating pion-nucleon coupling constant
$h_{\pi}^{1}$. Selected observables in two-nucleons systems are analyzed
in this framework with their dependence on these parameters being
expressed in a nearly model-independent way. 
\end{abstract}
\maketitle

\section{Introduction}

Study of the parity-violating (PV) nucleon-nucleon ($NN$) interaction,
$V^{\textrm{PV}}$, and its associated nuclear PV phenomena begins
with the report by Tanner~\cite{Tanner:1957} shortly after the parity
violation was confirmed experimentally in 1957. The first clear evidence
is found a decade after by observing a non-zero circular polarization
in the $\gamma$-decay of $^{181}\textrm{Ta}$~\cite{Lobashov:1967}.
Although quite a few PV observables have been measured later on in
various nuclear systems, ranging from simple two-body scattering to
heavy nuclear reaction, our current understanding of nuclear parity
violation is still far from complete (see, e.g., Refs.~\cite{Adelberger:1985ik,Haeberli:1995,Desplanques:1998ak,Ramsey-Musolf:2006dz}
for reviews of this field). The major difficulty is two-fold: not
only these experiments require high precision to discern the much
smaller PV signals, but also in theory, the non-perturbative character
of the quark-gluon dynamics makes a {}``first-principle'' formulation
of $V^{\textrm{PV}}$ as yet impossible. Despite the difficulties
and that the underlying theory, the $SU(2)\otimes U(1)$ gauge theory,
is well-established, the study of $V^{\textrm{PV}}$ is still valuable
for two main reasons. First, it is the only viable venue to observe
the neutral weak current interaction between two quarks at low energy.
Second, it supplies more information about the nucleon-nucleon ($NN$)
dynamics in addition to existing scattering data.

The phenomenological development of $V^{\textrm{PV}}$ proceeds in
a similar fashion as what has been achieved in the parity-conserving
(PC) $NN$ interaction---starting out from pure phase-shift analyses,
then parameterizations in meson exchange models, and finally to rigorous
effective field theory (EFT) formulation nowadays. It is first pointed
out by Danilov~\cite{Danilov:1965,Danilov:1971fh,Danilov:1972} that,
at low energy, $V^{\textrm{PV}}$ can be characterized by five $S\textrm{--}P$
scattering amplitudes: $\lambda_{s}^{pp}$, $\lambda_{s}^{np}$, and
$\lambda_{s}^{nn}$ for $^{1}S_{0}\textrm{--}^{3}P_{0}$ transitions,
$\lambda_{t}$ for $^{3}S_{1}\textrm{--}^{1}P_{1}$ transition, and
$\rho_{t}$ for $^{3}S_{1}\textrm{--}^{3}P_{1}$ transition. This
idea is generalized by Desplanques and Missimer~\cite{Desplanques:1978mt,Desplanques:1979ih}
to an effective version---through the Bethe-Goldstone equation---which
applies to many-body systems. On the other hand, formulations of $V^{\textrm{PV}}$
in terms of meson exchange models can be dated back to the seminal
works by Blin-Stoyle~\cite{Blin-Stoyle:1960a,Blin-Stoyle:1960b}
and Barton~\cite{Barton:1961eg}. The specific form involving one
$\pi$-, $\rho$-, and $\omega$-exchanges, $V_{\textrm{OME}}^{\textrm{PV}}$,
then becomes the standard in this field after Desplanques, Donoghue,
and Holstein (DDH) give their prediction of the six PV meson-nucleon
coupling constants, $h_{m}^{i}$'s ($m$ denotes the type of meson
and $i$ the isospin), based on a quark model calculation~\cite{Desplanques:1980hn}.\,%
\footnote{As the conventional nomenclature, the {}``DDH'' potential, could
be somewhat misleading, it is referred as the PV one-meson-exchange
(OME) potential, $V_{\textrm{OME}}^{\textrm{PV}}$, in this work.
We thank B. Desplanques for this clarification.%
} As explained in Ref.~\cite{Adelberger:1985ik}, $V_{\textrm{OME}}^{\textrm{PV}}$
has a close connection to the $S\textrm{--}P$ amplitude formulation,
$V_{S\textrm{--}P}^{\textrm{PV}}$, at low energy: The amplitude $\rho_{t}$
contains a long-range one-pion-exchange contribution, and the other
amplitudes, including the short-range part of $\rho_{t}$, are all
related to the vector-meson exchanges.

Most of the existing PV observables have been analyzed in the one-meson-exchange
(OME) framework. However, a consistent constraint of the PV coupling
constants is not realized yet. There are several reasons. On the experimental
side, many data have large errors so are not very constrictive; also,
these observables in terms of $h_{m}^{i}$'s are not independent enough
to allow a simultaneous determination of these six parameters. On
the theoretical side, several precise data involve many-body systems;
the reliabilities of these calculations are questionable. As one can
see from, e.g., Refs.~\cite{Haeberli:1995,Haxton:2001mi,Haxton:2001zq},
a two-dimensional constraint on the particular linear combinations
of the isoscalar and isovector couplings already shows some discrepancy.
Besides these possibilities, one might also wonder if the analysis
framework, i.e., $V_{\textrm{OME}}^{\textrm{PV}}$, could be the culprit.

To address the last question, Zhu \emph{}et al.~\cite{Zhu:2004vw}
recently reformulate $V^{\textrm{PV}}$ in the EFT framework to the
order of $Q$ ($Q$ is the momentum scale). This new framework comes
with two incarnations: one with pions fully integrated out, $V_{\slashed{\pi}}^{\textrm{PV}}$,
and the other with dynamical pions, $V_{\textrm{EFT}}^{\textrm{PV}}$.
The pionless version, thought to be suitable for low-energy cases
with $E_{\textrm{lab}}\lesssim100\,\textrm{MeV}$, only contains the
short-range (SR) interaction, $V_{1,\,\textrm{SR}}^{\textrm{PV}}$,
and it is specified by ten low-energy constants (LECs). However, as
Zhu \emph{}et al. argued semi-quantitatively, only five of them are
truly independent at low energy, and can be mapped to the five $S\textrm{--}P$
amplitudes. In the pionful version, dynamical pions generate two explicit
terms: the leading-order, long-range (LR) interaction, $V_{-1,\,\textrm{LR}}^{\textrm{PV}}$,
due to the one-pion exchange (OPE), and the subleading-order, medium-range
(MR) interaction, $V_{1,\,\textrm{MR}}^{\textrm{PV}}$, due to the
two-pion exchange (TPE). While the OPE part is also familiar in $V_{\textrm{OME}}^{\textrm{PV}}$,
the TPE part has never been systematically treated before. By considering
vertex corrections to the OPE term, the original formulation by Zhu
\textit{}\textit{\emph{et al}}\textit{.} contains an extra next-to-next-to-leading-order
interaction, $V_{1,\,\textrm{LR}}^{\textrm{PV}}$, whose operator
structure is thought to be different from others already being specified.
This term has recently been shown as redundant\,\cite{Liu:2006-u1},
so will be ignored in our discussion. Overall, in addition to the
$10$ LECs in the SR interaction, the pionful theory introduces, to
$\mathcal{O}(Q)$, two more parameters: one with the interaction ($h_{\pi}^{1}$)
and the other with the pion-exchange current ($\bar{c}_{\pi}$). An
important point to note is that, although $V_{1,\,\textrm{SR}}^{\textrm{PV}}$
takes the same form in both $V_{\mathrm{EFT}}^{\textrm{PV}}$ and
$V_{\slashed{\pi}}^{\textrm{PV}}$, the LECs in these two EFT frameworks
have different meanings: all the pion physics is included in LECs
for the pionless version; but it is singled out in the pionful version.

With the advance of experimental techniques showing promise of PV
measurements in few-body systems---where reliable theoretical calculations
are available---an extensive search program to re-analyze PV observables
is sketched in Ref.~\cite{Zhu:2004vw}. The key of this re-analysis
is to use the {}``hybrid'' EFT framework, which combines the state-of-the-art
wave functions (from phenomenological potential-model calculations)
and the most general form of $V^{\textrm{PV}}$(from EFT techniques).
The immediate goal is to find out whether a more consistent picture
of nuclear parity violation can be reached among few-body systems.
The long-term goal of including other precise measurements in many-body
systems certainly relies on the previous success.

This paper takes the first step dealing with two-nucleon systems at
low energy. The aim is to express the observables, both existing and
potentially possible, in terms the EFT parameters, and to serve as
a part of the database which the complete search program calls for.
The general formalism is introduced in Sec.~\ref{sec:formalism}.
The connection between $V^{\mathrm{PV}}$ (both $V_{\slashed{\pi}}^{\textrm{PV}}$
and $V_{\mathrm{EFT}}^{\mathrm{PV}}$) and the $S$--$P$ amplitudes
is studied in Sec.\,\ref{sec:SP amp.}. To facilitate a both realistic
and economic search program, special attention is on the quantitative
determination of a minimal set of parameters and its applicable energy
range. The observables of two-nucleon systems are discussed subsequently
in Sec.~\ref{sec:observables}, and a summary follows in Sec.~\ref{sec:summary}.

\section{Formalism~\label{sec:formalism}}

A fully consistent study of nuclear PV phenomena in the EFT framework
requires treating PC and PV interactions order by order on the same
footing. On the other hand, the {}``hybrid'' approach, which combines
the state-of-the-art wave functions from phenomenology and the general
operator structure from EFT, is shown to have quite some success.
In this work, we follow the latter approach as outlined in Ref.~\cite{Zhu:2004vw}.

\subsection{Parity-Conserving Potential and Wave Functions}

In the hybrid EFT framework, the unperturbed scattering and deuteron
(the binding energy $E_{\mathcal{D}}\cong2.22\,\textrm{MeV}$) wave
functions, $\ket{\psi}^{(\pm)}$ and $\ket{\psi}_{\mathcal{D}}$,
are obtained by solving the Lippmann-Schwinger and Schr\"{o}dinger
equations, respectively, 

\begin{align}
(H_{0}+V^{\textrm{PC}}\mp i\,\epsilon)\ket{\psi}^{(\pm)} & =E\ket{\psi}^{(\pm)}\,,\\
(H_{0}+V^{\textrm{PC}})\ket{\psi}_{\mathcal{D}} & =-E_{\mathcal{D}}\ket{\psi}_{\mathcal{D}}\,,\end{align}
with a chosen high-quality phenomenological potential as $V^{\textrm{PC}}$.
In this work, we use Argonne $v_{18}$ (AV18)~\cite{Wiringa:1995wb}
model exclusively. The model dependence of PV observables on strong
potentials has been extensively studied in Refs.~\cite{Carlson:2001ma,Schiavilla:2004wn};
for most cases, no strong deviation from AV18 is found.

Since the PV interaction is small, we treat it as a first-order perturbation.
The PV scattering amplitude, $\wt{M}$, is obtained by the first-order
distorted-wave Born approximation\begin{equation}
\wt{M}=^{(-)}\hspace{-0.15cm}\bra{\psi}V^{\textrm{PV}}\ket{\psi}^{(+)}\,.\end{equation}
The parity admixtures of the scattering and deuteron states, $\wt{\ket{\psi}}^{(\pm)}$
and $\wt{\ket{\psi}}_{\mathcal{D}}$, are obtained by solving the
inhomogeneous differential equations 

\begin{align}
(E-H_{0}-V^{\textrm{PC}})\wt{\ket{\psi}}^{(\pm)} & =V^{\textrm{PV}}\ket{\psi}^{(\pm)}\,,\\
(E_{\mathcal{D}}+H_{0}+V^{\textrm{PC}})\wt{\ket{\psi}}_{\mathcal{D}} & =-V^{\textrm{PV}}\ket{\psi}_{\mathcal{D}}\,,\end{align}
respectively, where the product of the PV potential and the unperturbed
wave function serves as the source term. We refer more technical details
regarding the partial wave expansion, phase shifts, and numerical
procedures to Refs.~\cite{Carlson:2001ma,Schiavilla:2004wn,Liu:2002bq},
but only mention a subtle point about the phase convention: We adopt,
exclusively, the Condon-Shortley phase convention; it is different
from the Biedenharn-Rose phase convention which contains an additional
phase $i^{L}$ for the partial wave of orbital angular momentum $L$.

\subsection{Parity-Violating Interaction in Pionless EFT}

In the pionless EFT, the PV interaction is entirely short-range and
takes the following form in the coordinate space~\cite{Zhu:2004vw} 

\begin{align}
V_{\slashed{\pi}}^{\textrm{PV}}(\bm r) & =V_{1,\textrm{SR}}^{\textrm{PV}}(\bm r)\nonumber \\
 & =\frac{2}{\Lambda_{\chi}^{3}}\,\left\{ \left[C_{1}+(C_{2}+C_{4})\,\tau_{+}^{z}+C_{3}\,\tau_{\cdot}+C_{5}\,\tau^{zz}\right]\,\bm\sigma_{-}\cdot\bm y_{m+}(\bm r)\right.\nonumber \\
 & \quad+\left[\wt{C}_{1}+(\wt{C}_{2}+\wt{C}_{4})\,\tau_{+}^{z}+\wt{C}_{3}\,\tau_{\cdot}+\wt{C}_{5}\,\tau^{zz}\right]\,\bm\sigma_{\times}\cdot\bm y_{m-}(\bm r)\nonumber \\
 & \quad+\left.(C_{2}-C_{4})\,\tau_{-}^{z}\,\bm\sigma_{+}\cdot\bm y_{m+}(\bm r)+\wt{C}_{6}\,\tau_{\times}^{z}\,\bm\sigma_{+}\cdot\bm y_{m-}(\bm r)\right\} \,,\label{eq:VPV-pionless}\end{align}
where $\Lambda_{\chi}$ is the scale of chiral symmetry breaking and
related to the pion decay constant $F_{\pi}$ by $\Lambda_{\chi}=4\,\pi\, F_{\pi}\approx1.161\,\textrm{GeV}$;
$\tau_{\cdot}\equiv\bm\tau_{1}\cdot\bm\tau_{2}$, $\tau_{\pm}^{z}\equiv(\tau_{1}^{z}\pm\tau_{2}^{z})/2$,
$\tau_{\times}^{z}\equiv i\,(\bm\tau_{1}\times\bm\tau_{2})^{z}/2$,
and $\tau^{zz}\equiv\left(3\,\tau_{1}^{z}\,\tau_{2}^{z}-\bm\tau_{1}\cdot\bm\tau_{2}\right)/(2\,\sqrt{6})$
are the isospin operators;~%
\footnote{The operators $\tau^{zz}$ and $\tau_{\times}^{z}$ we adopt are different
from Ref.~\cite{Zhu:2004vw}. Therefore, the LECs $C_{5}$, $\wt{C}_{5}$,
and $\wt{C}_{6}$ in our definition are greater than their counterparts
in Ref.~\cite{Zhu:2004vw} by factors of $-2\,\sqrt{6}$, $-2\,\sqrt{6}$,
and $2$, respectively. Also note that the notation of $C_{6}$ in
Ref.~\cite{Zhu:2004vw} is changed into $\wt{C}_{6}$ in this paper,
because it is associated with a $\bm y_{m-}$ type operator like other
$\wt{C}$'s.%
} $\bm\sigma_{\pm}\equiv\bm\sigma_{1}\pm\bm\sigma_{2}$ and $\bm\sigma_{\times}\equiv i\,\bm\sigma_{1}\times\bm\sigma_{2}$
are the spin operators. The spatial operator $\bm y_{m-(+)}(\bm r)$
is defined as the (anti-) commutator of $-i\,\bm\nabla$ with the
$\textrm{mass}^{2}$-weighted Yukawa function $f_{m}(r)$\begin{equation}
\bm y_{m\pm}(\bm r)=[-i\,\bm\nabla\,,\, f_{m}(r)]_{\pm}\equiv\left[-i\,\bm\nabla\,,\, m^{2}\,\frac{\textrm{e}^{-m\, r}}{4\,\pi\, r}\right]_{\pm}\,.\label{eq:dyfunc}\end{equation}
When $m\rightarrow\infty$, $f_{m}(r)\rightarrow\delta(r)/r^{2}$;
the potential thus takes a four-fermion contact form as expected. 

While using a Yukawa functional form for $f_{m}(r)$ leads to a similar
spatial behavior as the conventional $V_{\textrm{OME}}^{\textrm{PV}}$,
other choices---as long as they are realized in the context of EFT---are
also possible. For instance, taking into account the monopole form
factors at both the strong and weak vertices with a cutoff $\Lambda_{m}$,
one obtains a modified Yukawa function \begin{equation}
\bar{f}_{m}(r)=\frac{m^{2}}{4\,\pi\, r}\,\left\{ \textrm{e}^{-m\, r}-\textrm{e}^{-\Lambda_{m}\, r}\left[1+\frac{1}{2}\,(1-\frac{m^{2}}{\Lambda_{m}^{2}})\,\Lambda_{m}\, r\right]\right\} \,.\label{eq:Yukawa-modified}\end{equation}
At the $\Lambda_{m}\rightarrow\infty$ limit, $\bar{f}_{m}(r)$ recovers
the {}``bare'' Yukawa form $f_{m}(r)$. We note that in Refs.~\cite{Carlson:2001ma,Schiavilla:2004wn},
a recent and extensive OME analyses of two-body nuclear PV, the authors
adopt $\bar{f}_{m}(r)$ instead of the conventional choice $f_{m}(r)$. 

As $\wt{C}_{2}$ and $\wt{C}_{4}$ appear as a linear combination
$\wt{C}_{2}+\wt{C}_{4}$ in Eq.~(\ref{eq:VPV-pionless}), $V_{1,\mathrm{SR}}^{\textrm{PV}}$
contains $11-1=10$ LECs. After resolving the isospin and spin matrix
elements of all allowed two-nucleon states, the PV observables depend
on the following ten linear combinations of $C$'s and $\wt{C}$'s:

\begin{itemize}
\item $p\, p$ : $D_{v}^{pp}=C_{1}+C_{3}+C_{2}+C_{4}+C_{5}/\sqrt{6}$ and
$\wt{D}_{v}^{pp}=\wt{C}_{1}+\wt{C}_{3}+[\wt{C}_{2}+\wt{C}_{4}]+\wt{C}_{5}/\sqrt{6}$,
\item $n\, n$ : $D_{v}^{nn}=C_{1}+C_{3}-C_{2}-C_{4}+C_{5}/\sqrt{6}$ and
$\wt{D}_{v}^{nn}=\wt{C}_{1}+\wt{C}_{3}-[\wt{C}_{2}+\wt{C}_{4}]+\wt{C}_{5}/\sqrt{6}$,
\item $n\, p\,|_{T_{i}=T_{f}=1}$ : $D_{v}^{np}=C_{1}+C_{3}-2\, C_{5}/\sqrt{6}$
and $\wt{D}_{v}^{np}=\wt{C}_{1}+\wt{C}_{3}-2\,\wt{C}_{5}/\sqrt{6}$,
\item $n\, p\,|_{T_{i}=T_{f}=0}$ : $D_{u}=C_{1}-3\, C_{3}$ and $\wt{D}_{u}=\wt{C}_{1}-3\,\wt{C}_{3}$,
\item $n\, p\,|_{T_{i}\neq T_{f}}$ : $D_{w}=C_{2}-C_{4}$ and $\wt{D}_{w}=\wt{C}_{6}$.
\end{itemize}
In fact, $V_{1,\mathrm{SR}}^{\textrm{PV}}$ is tantamount to the $\rho$-
and $\omega$-sectors of $V_{\textrm{OME}}^{\textrm{PV}}$ if one
assumes i) $m_{\rho}=m_{\omega}=m$ and ii) the following relations
between $\wt{C}$'s and $C$'s: \begin{subequations}\begin{align}
\frac{\wt{C}_{1}}{C_{1}}=\frac{\wt{C}_{2}}{C_{2}} & =1+\chi_{\omega}\,,\label{eq:w-collapse}\\
\frac{\wt{C}_{3}}{C_{3}}=\frac{\wt{C}_{4}}{C_{4}}=\frac{\wt{C}_{5}}{C_{5}} & =1+\chi_{\rho}\,,\label{eq:r-collapse}\end{align}
\end{subequations} where $\chi_{\omega}$ and $\chi_{\rho}$ are
the isoscalar and isovector strong tensor couplings, respectively.
The remaining $11-5=6$ independent LECs in EFT then have a one-to-one
mapping to the PV heavy-meson-nucleon coupling constants as \begin{subequations}\begin{align}
(C_{1},\, C_{2}) & \rightarrow-\frac{g_{\omega}}{2}\,(h_{\omega}^{0},\, h_{\omega}^{1})\,\frac{\Lambda_{\chi}^{3}}{\mN\, m_{\omega}^{2}}\,,\label{eq:wtoDDH}\\
(C_{3},\, C_{4},\, C_{5},\,\wt{C}_{6}) & \rightarrow-\frac{g_{\rho}}{2}\,(h_{\rho}^{0},\, h_{\rho}^{1},\, h_{\rho}^{2},\, h_{\rho}^{1'})\,\frac{\Lambda_{\chi}^{3}}{\mN\, m_{\rho}^{2}}\,,\label{eq:rtoDDH}\end{align}
\end{subequations}where $g_{x}$ denotes the strong $x$-meson-nucleon
coupling constant. Note that in analyses based on $V_{\textrm{OME}}^{\textrm{PV}}$,
the $h_{\rho}^{1'}$ part is usually ignored because it has the same
operator structure as the LR OPE interaction, i.e.\emph{,} the $h_{\pi}^{1}$
part, but a very small predicted value for $h_{\rho}^{1'}$. One might
be tempted to adopt a different Yukawa mass $m=m_{\pi}$ for the $\wt{C}_{6}$
term so that $V_{1,\mathrm{SR}}^{\textrm{PV}}$ bears even more similarity
to $V_{\textrm{OME}}^{\textrm{PV}}$. However, this is undesirable
from the EFT point of view because two quite different length scales,
$1/m_{\rho}$ and $1/m_{\pi}$, both show up in the {}``short-range''
interaction.

\subsection{Parity-Violating Interaction in Pionful EFT}

When pions are treated explicitly, the EFT PV interaction, as formulated
in Ref.\,\cite{Zhu:2004vw}, contains three parts\,%
\footnote{As mentioned in the introduction, we ignore the higher-order LR term
$V_{1,\mathrm{LR}}^{\mathrm{PV}}$ in Ref.\,\cite{Zhu:2004vw}, since
it is shown to be redundant\,\cite{Liu:2006-u1}.%
} \begin{equation}
V_{\textrm{EFT}}^{\textrm{PV}}(\bm r)=V_{-1,\textrm{LR}}^{\textrm{PV}}(\bm r)+V_{1,\textrm{MR}}^{\textrm{PV}}(\bm r)+V_{1,\textrm{SR}}^{\textrm{PV}}(\bm r)\,.\label{eq:VPV-EFT}\end{equation}

The leading term $V_{-1,\textrm{LR}}^{\textrm{PV}}$ is the familiar
PV OPE potential\begin{equation}
V_{-1,\textrm{LR}}^{\textrm{PV}}(\bm r)=\frac{2}{\Lambda_{\chi}^{3}}\,\wt{C}_{6}^{\pi}\,\tau_{\times}^{z}\,\bm\sigma_{+}\cdot\bm y_{\pi-}(\bm r)\,,\label{eq:VPV-OPE}\end{equation}
with\begin{equation}
\wt{C}_{6}^{\pi}=\frac{h_{\pi}^{1}\, g_{A}}{2\,\sqrt{2}}\,\frac{\Lambda_{\chi}^{3}}{F_{\pi}\, m_{\pi}^{2}}=\frac{h_{\pi}^{1}\, g_{\pi}}{2\,\sqrt{2}}\,\frac{\Lambda_{\chi}^{3}}{\mN\, m_{\pi}^{2}}\,,\label{eq:Cpi6}\end{equation}
where $g_{A}=1.27$ is the nucleon axial vector coupling constant
and $m_{\pi}=139.57\,\textrm{MeV}$; and the second equality follows
from the Goldberger-Trieman relation.

\begin{figure}
{\centering\begin{tabular}{cc}
\includegraphics{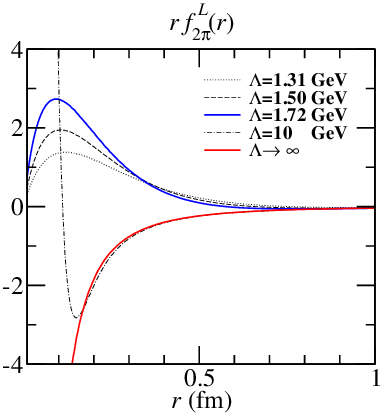}&
\includegraphics{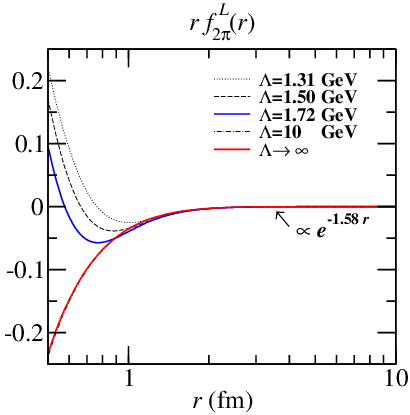}\tabularnewline
\includegraphics{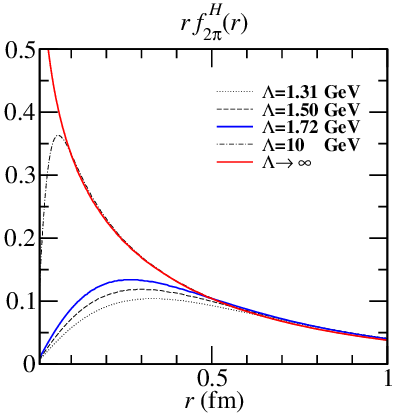}&
\includegraphics{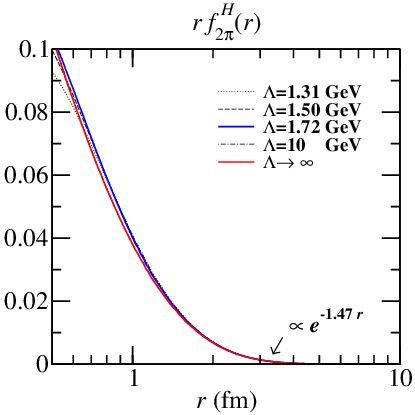}\tabularnewline
\end{tabular}}

\caption{The $r$-weighted Yukawa-like radial functions $r\, f_{2\pi}^{L}(r)$
and $r\, f_{2\pi}^{H}(r)$ in the two-pion-exchange potential.\,\label{fig:fTPE}}
\end{figure}

The subleading MR interaction is due to TPE and has the form\,%
\footnote{Some typographical errors in Eq.\,(121) of Ref.\,\cite{Zhu:2004vw}
have been fixed in order to produce Eqs. (\ref{eq:C2pi2}, \ref{eq:C2pi6});
see also Ref.\,\cite{Hyun:2006mp}. We thank B. Desplanques and Zhu
\textit{\emph{et al}}. for pointing this out.%
}\begin{align}
V_{1,\textrm{MR}}^{\textrm{PV}}(\bm r)= & \frac{2}{\Lambda_{\chi}^{3}}\,\Big\{\wt{C}_{2}^{2\pi}\,\tau_{+}^{z}\,\bm\sigma_{\times}\cdot\bm y_{2\pi}^{L}(\bm r)\nonumber \\
 & +\wt{C}_{6}^{2\pi}\,\tau_{\times}^{z}\,\bm\sigma_{+}\cdot\left[\left(1-1/(3\, g_{A}^{2})\right)\,\bm y_{2\pi}^{L}(\bm r)-1/3\,\bm y_{2\pi}^{H}(\bm r)\right]\Big\}\,,\label{eq:VPV-TPE}\end{align}
with\begin{align}
\wt{C}_{2}^{2\pi} & =-4\,\sqrt{2}\,\pi\, g_{A}^{3}\, h_{\pi}^{1}\,,\label{eq:C2pi2}\\
\wt{C}_{6}^{2\pi} & =3\,\sqrt{2}\,\pi\, g_{A}^{3}\, h_{\pi}^{1}\,.\label{eq:C2pi6}\end{align}
The Yukawa-like radial functions $f_{2\pi}^{L}(r)$ and $f_{2\pi}^{H}(r)$,
for generating $\bm y_{2\pi}^{L}(\bm r)$ and $\bm y_{2\pi}^{H}(\bm r)$
via Eq.\,(\ref{eq:dyfunc}), are obtained from the Fourier transforms
of \begin{align}
L(q) & =\frac{\sqrt{4\, m_{\pi}^{2}+\bm q^{2}}}{|\bm q|}\,\ln\left(\frac{\sqrt{4\, m_{\pi}^{2}+\bm q^{2}}+|\bm q|}{2\, m_{\pi}}\right)\,,\\
H(q) & =\frac{4\, m_{\pi}^{2}}{4\, m_{\pi}^{2}+\bm q^{2}}\, L(q)\,,\end{align}
respectively. In Fig.\,\ref{fig:fTPE}, the plots of $r\, f_{2\pi}^{L}(r)$
and $r\, f_{2\pi}^{H}(r)$ are shown with several dipole cutoff factors
$(\Lambda^{2}-4m_{\pi}^{2})^{2}/(\Lambda^{2}+q^{2})^{2}$---including
the bare case, \textit{\emph{i.e.}}, $\Lambda\rightarrow\infty$---introduced
in the Fourier transforms. As one clearly sees, the short distance
behaviors are quite cutoff-sensitive, especially for $f_{2\pi}^{L}(r)$
since $L(q)$ diverges logarithmically as $\ln q/m_{\pi}$. On the
other hand, the long-range tails, roughly decrease like $e^{-1.58\, r}$
and $e^{-1.47\, r}$, track well with $e^{-2\, m_{\pi}\, r}=e^{-1.41\, r}$.

The way we define $\wt{C}_{6}^{\pi}$, $\wt{C}_{2}^{2\pi}$ and $\wt{C}_{6}^{2\pi}$
is handy for the bookkeeping purpose; this gives $V_{-1,\textrm{LR}}^{\textrm{PV}}$
and $V_{1,\textrm{MR}}^{\textrm{PV}}$ the same formal structure as
the corresponding parts---as hinted by the subscripts---in $V_{1,\textrm{SR}}^{\textrm{PV}}$
(also, all the Yukawa functions $f_{m}(r),\, f_{\pi}(r),\, f_{2\pi}^{L}(r),\,\mbox{and}\, f_{2\pi}^{H}(r)$
have the same limit when $m,\, m_{\pi}\rightarrow\infty$). However,
this does not imply these pion PV constants are comparable in magnitude
with LECs $C$'s and $\wt{C}$'s. Comparing Eqs.\,(\ref{eq:Cpi6},
\ref{eq:C2pi2}, \ref{eq:C2pi6}), one sees $\wt{C}_{2,6}^{2\pi}$
smaller than $\wt{C}_{6}^{\pi}$ roughly by an order of magnitude.
By Eqs.\,(\ref{eq:wtoDDH}, \ref{eq:rtoDDH}) and assuming all the
$\pi$-, $\rho$-, and $\omega$-coupling constants approximately
the same, one estimates $C$'s smaller than $\wt{C}_{6}^{\pi}$ roughly
by a factor of $m_{\rho}^{2}/m_{\pi}^{2}\sim30$; for $\wt{C}$'s,
due to the tensor couplings, Eqs.\,(\ref{eq:w-collapse}, \ref{eq:r-collapse}),
the suppression can be less. If the above assumptions are not too
far off, we can roughly conclude that $\wt{C}_{2,6}^{2\pi}$ and the
LECs, $C$'s and $\wt{C}$'s, are of the same order, and all of them
smaller than $\wt{C}_{6}^{\pi}$ by an order of magnitude. This observation
is consistent with the power counting scheme that both the TPE and
SR terms are of the same higher order than the OPE one. But, more
definitive answer should still be sought from experiments.

We also emphasize that the TPE term $V_{1,\textrm{MR}}^{\textrm{PV}}$,
Eq.\,(\ref{eq:VPV-TPE}), being used here is not the full PV TPE
potential. According to Ref.\,\cite{Zhu:2004vw}, it only contains
the singular part of the TPE, and the rest regular terms are still
effectively included in the short-range interaction. In this sense,
$V_{\textrm{EFT}}^{\textrm{PV}}$ in fact depends on the chosen regularization
scheme. However, as long as the most general operator structure is
maintained and one does not try to fit the data over a large energy
range, a consistent analysis should be scheme-independent---this will
become clear in the next section.

\subsection{Setup and Parameters}

In the following sections, various PV observables in two-nucleon systems
will be analyzed by $V_{\mathrm{EFT}}^{\textrm{PV}}$. While the full
results are realized in the pionful EFT framework, the ones correspond
to the pionless EFT framework can be easily read by simply retaining
only the part from $V_{1,\,\textrm{SR}}^{\textrm{PV}}$---with the
notion that the LECs in this case effectively include all the pion
contributions. Although the choice of the Yukawa mass parameter in
$V_{1,\,\textrm{SR}}^{\textrm{PV}}$ is arbitrary, we use the $\rho$
meson mass, $m=m_{\rho}=771.1\,\textrm{MeV}$, since it has an easy
connection to the meson-exchange picture. For the convenience of presentation,
these calculations will be referred as the {}``bare'' calculations,
because all Yukawa functions in $V_{\mathrm{EFT}}^{\textrm{PV}}$
are not modified by any form factor. 

The above results will be checked against existing calculations in
the $V_{\textrm{OME}}^{\textrm{PV}}$ framework. This is done by applying
the relations Eqs.~(\ref{eq:w-collapse}, \ref{eq:r-collapse}, \ref{eq:rtoDDH},
\ref{eq:wtoDDH}, and \ref{eq:Cpi6}) to $V_{\mathrm{EFT}}^{\textrm{PV}}$
and ignoring all the TPE contribution; from now on, we call this procedure
OME-mapping (OME-m). For numerical estimates, the strong parameters
are taken from Ref.~\cite{Adelberger:1985ik} and the weak ones are
set to be the {}``best-guess'' values of DDH~\cite{Desplanques:1980hn}.
This set is labeled as {}``DDH-best'' in Tabs.~\ref{cap:strong-parameters}
and \ref{cap:weak-parameters}. 

In order to compare with Refs.~\cite{Carlson:2001ma,Schiavilla:2004wn},
as pointed out previously, one has to use the monopole-modified Yukawa
functions instead. For this matter, we perform a parallel set of calculations
using $\bar{f}_{m}(r)$ with the cutoff parameters chosen to be $\Lambda_{\rho}=\Lambda_{\omega}=1.50\,\textrm{GeV}$
and $\Lambda_{\pi}=\Lambda_{2\pi}=1.72\,\textrm{GeV}$. These calculations
will be referred as the {}``mod'' calculations. While using $\bar{f}_{\rho,\omega}(r)$
in $V_{1,\mathrm{SR}}^{\mathrm{PV}}$ does not contradict the EFT
framework, using $\bar{f}_{\pi}(r)$ and $\bar{f}_{2\pi}^{L,H}(r)$
does not seem fully consistent with EFT. For the OPE part, this is
less a problem because form factors only suppress the LR OPE slightly.
On the other hand, the validity of adding form factors to the TPE
part in EFT needs further justification. Thus, our calculation for
this TPE part should only be understood as showing a qualitative feature
in cases where form factors are built in. For numerical results in
this set of calculations, the strong parameters are taken from the
Bonn model~\cite{Machleidt:2000ge} and the weak ones are the fitted
results of Refs.~\cite{Carlson:2001ma,Schiavilla:2004wn}. This set
is labeled as {}``DDH-adj.'' in Tabs.~\ref{cap:strong-parameters}
and \ref{cap:weak-parameters}.

\begin{table}

\caption{Sets of strong parameters used for OME-mapping.~\label{cap:strong-parameters}}

\begin{ruledtabular}\begin{tabular}{cccccc}
&
$g_{\pi}$&
$g_{\rho}$&
$g_{\omega}$&
$\kappa_{\rho}$&
$\kappa_{\omega}$\tabularnewline
\hline 
DDH-best~\cite{Adelberger:1985ik}&
$13.45$&
$2.79$&
$8.37$&
$3.70$&
$-0.12$\tabularnewline
DDH-adj.~\cite{Schiavilla:2004wn}&
$13.22$&
$3.25$&
$15.85$&
$6.10$&
$0.0$\tabularnewline
\end{tabular}\end{ruledtabular}
\end{table}
\begin{table}

\caption{Sets of weak parameters used for OME-mapping (in units of $10^{-7}$).
Note that for $p\, p$ systems, $h_{\rho}^{0}+h_{\rho}^{1}+h_{\rho}^{2}/\sqrt{6}=-22.3$,
instead of $-24.8$ as shown in this table, will be used~\cite{Carlson:2001ma}.~\label{cap:weak-parameters}}

\begin{ruledtabular}\begin{tabular}{cccccccc}
&
$h_{\pi}^{1}$&
$h_{\rho}^{0}$&
$h_{\rho}^{1}$&
$h_{\rho}^{2}$&
$h_{\omega}^{0}$&
$h_{\omega}^{1}$&
$h_{\rho}^{1'}$\tabularnewline
\hline 
DDH-best~\cite{Desplanques:1980hn}&
$4.56$&
$-11.4$&
$-0.19$&
$-9.50$&
$-1.90$&
$-1.14$&
$0.00$\tabularnewline
DDH-adj.~\cite{Carlson:2001ma,Schiavilla:2004wn}&
$4.56$&
$-16.4$&
$-2.77$&
$-13.7$&
$+3.23$&
$+1.94$&
$0.00$\tabularnewline
\end{tabular}\end{ruledtabular}
\end{table}

\section{$S$--$P$ Amplitudes and Danilov Parameters\,\label{sec:SP amp.}}

A PV potential with 11 (10 LECs plus $h_{\pi}^{1}$) undetermined
parameters certainly poses a formidable challenge---how can we gather
sufficient data and do reliable theoretical analyses of them? A substantial
reduction of the LECs is proposed in Refs.\,\cite{Zhu:2004vw} by
building the connection between the EFT framework and the $S$--$P$
amplitude analysis, which is pioneered by by Danilov~\cite{Danilov:1965,Danilov:1971fh,Danilov:1972},
and extended by Desplanques and Missimer later on~\cite{Desplanques:1978mt,Desplanques:1979ih,Desplanques:1980}.
The main idea of this reduction goes like the following.

For low-energy PV phenomena in which only the $S$--$P$ mixings contribute
substantially, the observables can be expressed by five independent
PV scattering amplitudes: $v_{pp,nn,np}(^{1}S_{0}\rightarrow^{3}P_{0})$,
$u(^{3}S_{1}\rightarrow^{1}P_{1})$, and $w(^{3}S_{1}\rightarrow^{3}P_{1})$.
From the last section, we know each amplitude due to the $V_{1,\mathrm{SR}}^{\mathrm{PV}}$
part is a linear combination of the corresponding $D$ and $\wt{D}$
with the coefficients being determined by the matrix elements of $\bm y_{m+}$
and $\bm y_{m-}$, respectively. An important observation comes from
that the matrix elements $\langle\bm y_{m+}\rangle$ and $\langle\bm y_{m-}\rangle$
are equal in the zero range approximation (ZRA) with the absence of
the $NN$ repulsive hard core. This causes $D$ and $\wt{D}$ always
appear in a $(D+\wt{D})$ combination and work actually like one energy-independent
LEC. Therefore, the number of LECs can be reduced to 5 which corresponds
to the number of independent $S$--$P$ amplitudes. While this sounds
an attractive idea, however, neither ZRA nor the absence of $NN$
hard core are physically realized. In order to put this reduction
scheme of LECs along with the whole search program proposed in Ref.\,\cite{Zhu:2004vw}
on a firmer ground, we try to answer a series of questions which have
not been addressed: 

\begin{enumerate}
\item For realistic cases, this 10-to-5 reduction can still work as long
as the matrix elements of $\bm y_{m+}$ and $\bm y_{m-}$ have (almost)
the same energy dependence, \textit{\emph{i.e.}}, the condition\begin{equation}
\frac{\langle\bm y_{m+}\rangle}{\langle\bm y_{m-}\rangle}\equiv R(E)\cong R\,,\label{eq:scaling}\end{equation}
is satisfied. Therefore, we try to determine $R$ and its range of
constancy over $E$ for each $S$--$P$ amplitude.
\item When higher partial waves become important, the $S$--$P$ analyses
are no longer valid. Thus, it is necessary to determine the maximum
energy for this proposed search program to work.
\item In Ref.\,~\cite{Zhu:2004vw}, it is also proposed that for $E_{\mathrm{lab}}\lesssim100\,\mbox{MeV}$,
the pionless EFT should work. However, this requires the contributions
from $V_{-1,\mathrm{LR}}^{\mathrm{PV}}$ and $V_{1,\mathrm{MR}}^{\mathrm{PV}}$
can be effectively included in $V_{1,\mathrm{SR}}^{\mathrm{PV}}$.
We try to justify whether this condition can be met.
\item At the end of this section, we determine the zero-energy $S$--$P$
amplitudes, the so-called Danilov parameters, in terms of the PV parameters
in $V_{\mathrm{EFT}}^{\mathrm{PV}}$. They will be the actual parameters
used to express the PV observables in the next section. 
\end{enumerate}
According to the definitions by Desplanques and Missimer\,\cite{Desplanques:1978mt},
the $S$--$P$ amplitudes are calculated by the following formulae:\begin{subequations}
\begin{align}
v & =\frac{\mN}{i\, p}\,\frac{\langle^{3}P_{0}|V_{\mathrm{EFT}}^{\mathrm{PV}}|^{1}S_{0}\rangle}{\langle^{3}P_{0}|\bm\sigma_{-}\cdot\hat{r}|^{1}S_{0}\rangle}\,\frac{e^{\eta\,\pi}}{|\Gamma(1+i\,\eta)|\,|\Gamma(2+i\,\eta)|}\nonumber \\
 & =v^{-}\,\wt{D}_{v}+v^{+}\, D_{v}+v^{2\pi}\,\wt{C}_{2}^{2\pi}\,,\\
u & =\frac{-\mN}{i\, p}\,\frac{\langle^{1}P_{1}|V_{\mathrm{EFT}}^{\mathrm{PV}}|^{3}S_{1}\rangle}{\langle^{1}P_{1}|\bm\sigma_{-}\cdot\hat{r}|^{3}S_{1}\rangle}\nonumber \\
 & =u^{-}\,\wt{D}_{u}+u^{+}\, D_{u}\,,\\
w & =\frac{-\mN}{i\, p}\,\frac{\langle^{3}P_{1}|V_{\mathrm{EFT}}^{\mathrm{PV}}|^{3}S_{1}\rangle}{\langle^{3}P_{1}|\bm\sigma_{+}\cdot\hat{r}|^{3}S_{1}\rangle}\nonumber \\
 & =w^{-}\,\wt{D}_{w}+w^{+}\, D_{w}+w^{\pi}\,\wt{C}_{6}^{\pi}+w^{2\pi}\,\wt{C}_{6}^{2\pi}\,,\end{align}
\end{subequations}where all the amplitudes are functions of energy,
$p$ is the two-nucleon relative momentum, and the factors are chosen
to reproduce the normalization and limiting behaviors of Refs.\,
\cite{Desplanques:1978mt,Desplanques:1998ak}. Note that for the $v$
amplitude, an extra factor, which is $1$ when the Sommerfeld number
$\eta=0$, is introduced in order to completely subtract the Coulomb
effect for $p\, p$ scattering at threshold. 

Since the total cross section is proportional to the imaginary part
of the forward scattering amplitude, we concentrate on $\mbox{Im[}v,u,w\mbox{]}$.

\begin{figure}
\begin{tabular}{ccc}
\includegraphics[scale=0.7]{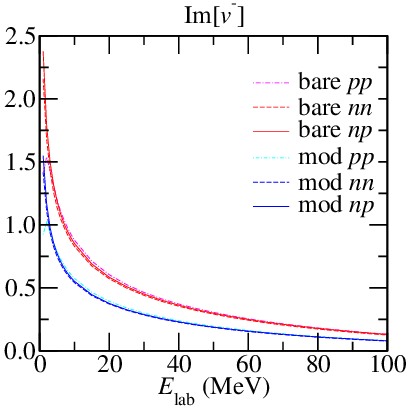}&
\includegraphics[scale=0.7]{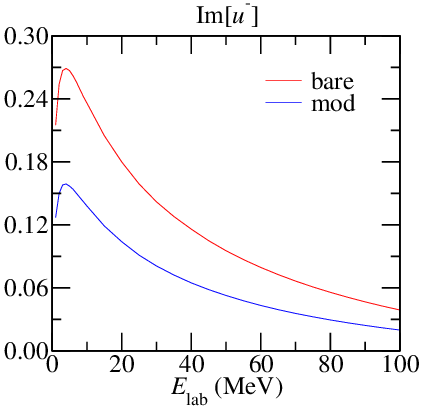}&
\includegraphics[scale=0.7]{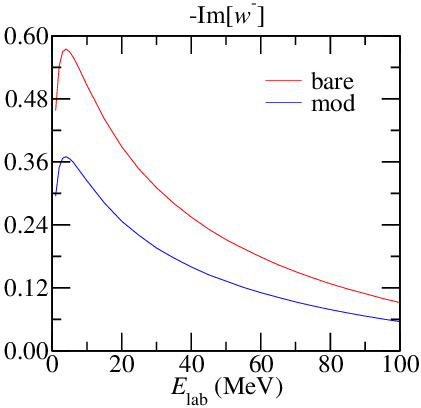}\tabularnewline
\includegraphics[scale=0.7]{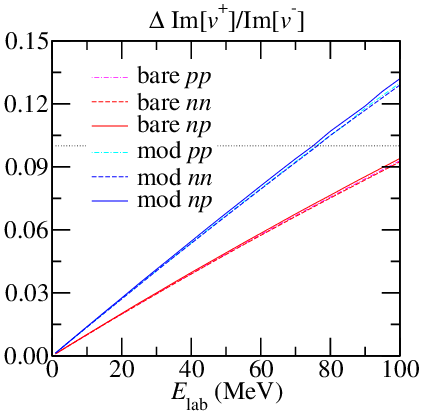}&
\includegraphics[scale=0.7]{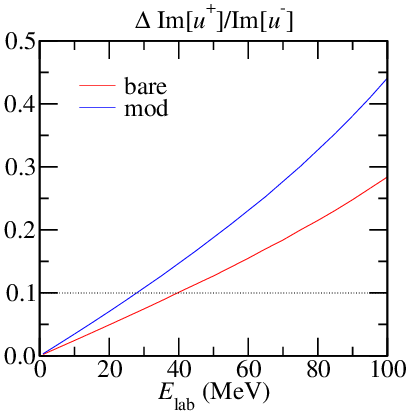}&
\includegraphics[scale=0.7]{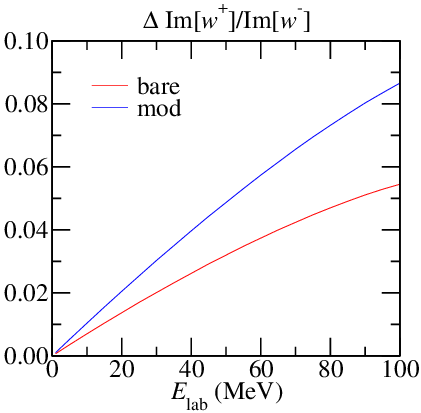}\tabularnewline
\end{tabular}

\caption{Top panels: the energy dependences of the $x^{-}$ type $S$--$P$
amplitudes. Bottom panels: the percentage deviations of $x^{+}/x^{-}$
from their zero-energy values $x^{+}(0)/x^{-}(0)$, where the dotted
lines mark the $10\%$ level.\,\label{fig:10-to-5}}
\end{figure}

The top panels of Fig.\,\ref{fig:10-to-5} show the energy dependences
of $x^{-}$ ($x\in v_{pp,nn,np},\, u,\, w$) type $S$--$P$ amplitudes,
proportional to $\langle\bm y_{m-}\rangle$, up to $E_{\mathrm{lab}}=100\,\mbox{MeV}$.
Because the form factors suppress the short-distance contributions,
the results from the {}``mod'' calculations are consistently smaller
than the {}``bare'' ones. Another noticeable feature is the plots
of $v_{pp}$, $v_{nn}$, and $v_{np}$ overlap, and the tiny difference
is mostly due to the small charge-dependent interaction built in AV18.

The bottom panels of Fig.\,\ref{fig:10-to-5} show the percentage
deviations of $x^{+}/x^{-}$ from their zero-energy values $x^{+}(0)/x^{-}(0)$,
\textit{\emph{i.e.}}, \begin{equation}
\Delta\, x^{+}/x^{-}\equiv\frac{x^{+}/x^{-}-x^{+}(0)/x^{-}(0)}{x^{+}(0)/x^{-}(0)}\,.\end{equation}
In case the energy dependences of $x^{+}$ and $x^{-}$ keep the same,
$\Delta\, x^{+}/x^{-}$ remains zero. Therefore, this quantity is
a measure of the departure from the perfect scaling, Eq.\,(\ref{eq:scaling}),
which a strict 10-to-5 reduction scheme requires. As these plots show,
the deviations all grow with energy in the positive direction. For
the $v$ and $w$ amplitudes, the scaling actually works very well---though
not perfect---up to $E_{\mathrm{lab}}=100\,\mbox{MeV}$ where the
deviations are still less than $10\%$ for the {}``bare'' case.
For the $u$ amplitude, however, the $10\%$ tolerance for scaling
deviation can only allow $E_{\mathrm{lab}}$ go as high as $40\,\mbox{MeV}$.
It might come as a surprise why the $u$ and $w$ amplitudes have
such different behaviors, since both involve the same $^{3}S_{1}$
wave and the $^{1}P_{1}$ wave (for $u$) does not differ from the
$^{3}P_{1}$ one (for $w$) dramatically. The answer is due to the
tensor force, by which a true distorted $^{3}S_{1}$-wave acquires
some $D$-wave component. From the ratios\begin{align}
\langle^{1}P_{1}|\bm\sigma_{-}\cdot\hat{r}|^{3}S_{1},L=2\rangle/\langle^{1}P_{1}|\bm\sigma_{-}\cdot\hat{r}|^{3}S_{1},L=0\rangle & =-\sqrt{2}\,,\label{eq:spinang-u}\\
\langle^{3}P_{1}|\bm\sigma_{+}\cdot\hat{r}|^{3}S_{1},L=2\rangle/\langle^{3}P_{1}|\bm\sigma_{+}\cdot\hat{r}|^{3}S_{1},L=0\rangle & =1/\sqrt{2}\,,\label{eq:spinang-w}\end{align}
one learns that the $D$-wave component is more enhanced in the $u$
than the $w$ amplitude, and this is the main cause of the difference.
A mock calculation by pretending the ratio in Eq.\,(\ref{eq:spinang-u})
to be $1/\sqrt{2}$ as in Eq.\,(\ref{eq:spinang-w}) verifies that
the scaling of $u^{+}/u^{-}$ would then be similar to $w^{+}/w^{-}$.
The {}``mod'' calculations generally show larger deviations, so
the ranges within which the scalings work have to be reduced. This
can be understood from that the difference of $\bm y_{m+}$ and $\bm y_{m-}$
is a term involving the gradient on the wave function. As the form
factors suppress the short-range contribution, the longer-range part
of the wave function which has a larger gradient thus gets a bigger
weight; this leads to an enhancement of the deviation. 

\begin{figure}
\begin{tabular}{ccc}
\includegraphics[scale=0.7]{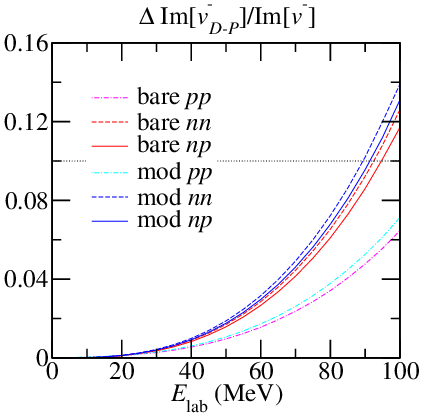}&
\includegraphics[scale=0.7]{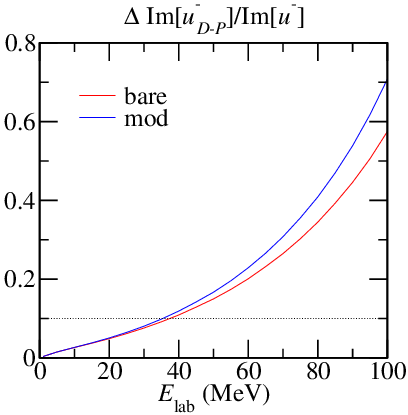}&
\includegraphics[scale=0.7]{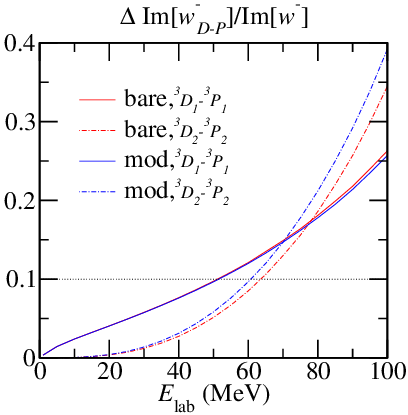}\tabularnewline
\end{tabular}

\caption{The percentage corrections induced by the $D$--$P$ transitions
to the $S$--$P$ ones for the $x^{-}$ type amplitudes, where the
dotted lines mark the $10\%$ level.\,\label{fig:DP-to-SP}}
\end{figure}

The importance of the $D$--$P$ transitions is illustrated in Fig.\,\ref{fig:DP-to-SP},
where their percentage corrections to the $S$--$P$ transitions are
shown for the $x^{-}$ type amplitudes. Given a $10\%$ tolerance
for corrections, one can conclude that the $S$--$P$ transitions
dominate up to $E_{\mathrm{lab}}=90\,\mbox{MeV}$ for the $v$ type
amplitudes (it can be higher for the $p\, p$ case because of the
Coulomb repulsion) and up to $E_{\mathrm{lab}}=40\,\mbox{MeV}$ for
both the $u$ and the $w$ types (note that the $w$ type contains
two contributions from $^{3}D_{1}$--$^{3}P_{1}$ and $^{3}D_{2}$--$^{3}P_{2}$).
The {}``mod'' calculations do not change these conclusions too much.
It should be stressed that these $D$--$P$ amplitudes have quite
different energy dependences compared to the $S$--$P$ ones. Therefore,
although the scaling behaviors found above for the $S$--$P$ amplitudes
can apply to higher energies, the 10-to-5 reduction should be limited
by the prerequisite of the $S$--$P$ dominance.

\begin{figure}
\begin{tabular}{ccc}
\includegraphics[scale=0.7]{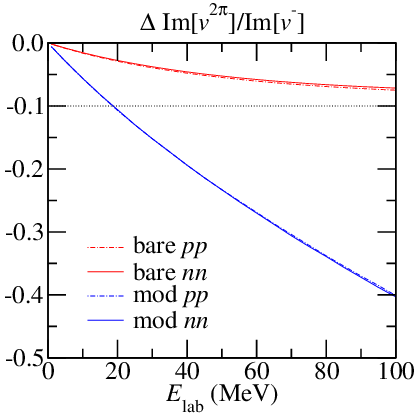}&
\includegraphics[scale=0.7]{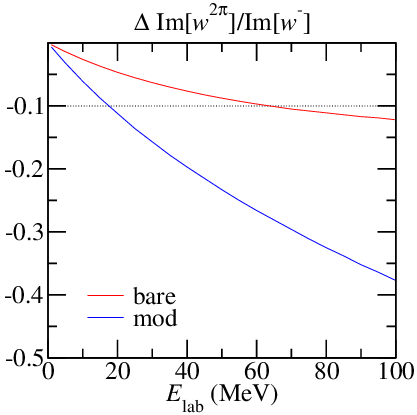}&
\includegraphics[scale=0.7]{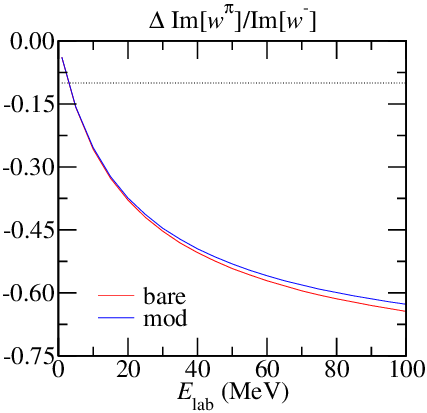}\tabularnewline
\end{tabular}

\caption{The percentage deviations of $x^{\pi,2\pi}/x^{-}$ from their zero-energy
values $x^{\pi,2\pi}(0)/x^{-}(0)$, where the dotted lines mark the
$10\%$ level.\,\label{fig:pi-to-SR}}
\end{figure}

In Fig.\,\ref{fig:pi-to-SR}, whether the OPE and TPE contributions
can be effectively integrated out and lead to a purely pionless short-range
potential is studied by examining the scaling behaviors of one- and
two-pion amplitudes $x^{\pi,2\pi}$ with respect to $x^{-}$. As one
can see, all the deviations increase with energy in the negative direction.
In the {}``bare'' calculations, the TPE amplitudes scale very well
with the SR one. Allowing a $10\%$ deviation, the scaling works as
$E_{\mathrm{lab}}$ reaches up to more than $100\,\mbox{MeV}$ for
the $v$ amplitudes and $60\,\mbox{MeV}$ for the $w$ one; both limits
are above the $S$--$P$-dominant region. On the other hand, the scaling
works quite poorly---only up to $E_{\mathrm{lab}}=20\,\mbox{MeV}$---in
the {}``mod'' calculations. This can be easily seen from Fig.\,\ref{fig:fTPE},
where the modified two-pion Yukawa-like functions differs substantially
from the bare ones at short distances. Since we have already mentioned
the inconsistency of adding form factors to the TPE potential in such
an \emph{}\textit{\emph{ad hoc}} fashion in EFT, the curves labeled
by {}``mod'' for the TPE part should not be taken too seriously.

The most noteworthy information in Fig.\,\ref{fig:pi-to-SR} comes
from the observation that it is almost impossible for the OPE amplitude
to scale with the SR one, even within a modest energy range, say $10\,\mbox{MeV}$
or so. From threshold to $40\,\mbox{MeV}$, in which the $S$--$P$
dominance holds for the $w$ amplitude, the scaling deviation increases
to $50\%$. Thus, this re-confirms the old wisdom that it takes two
parameters---one for the SR and the other for the LR terms---to characterize
the physics of the $^{3}S_{1}$--$^{3}P_{1}$ transition\,\cite{Adelberger:1985ik}.
In this sense, the applicability of the pionless EFT to nuclear PV
is extremely limited, only within a very narrow energy range near
threshold. 

\begin{figure}
\begin{tabular}{ccc}
\includegraphics[scale=0.7]{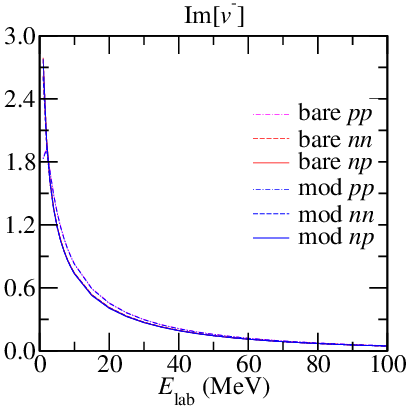}&
\includegraphics[scale=0.7]{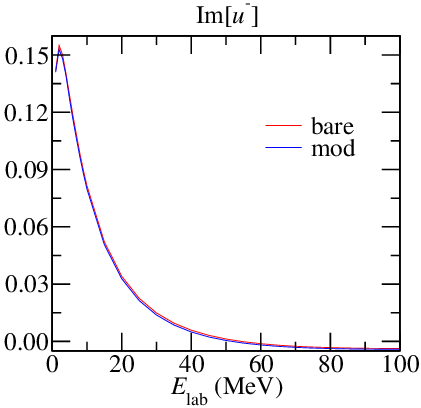}&
\includegraphics[scale=0.7]{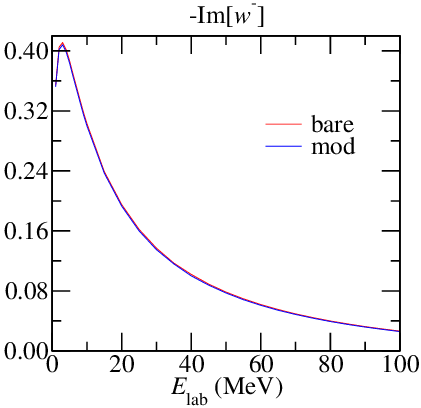}\tabularnewline
\includegraphics[scale=0.7]{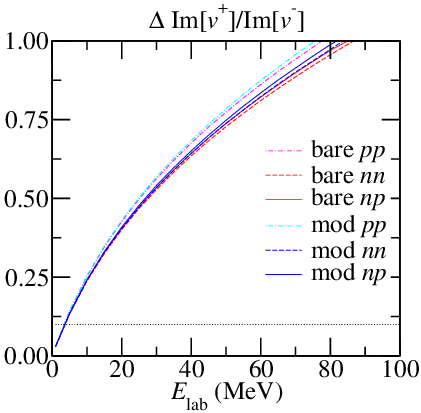}&
\includegraphics[scale=0.7]{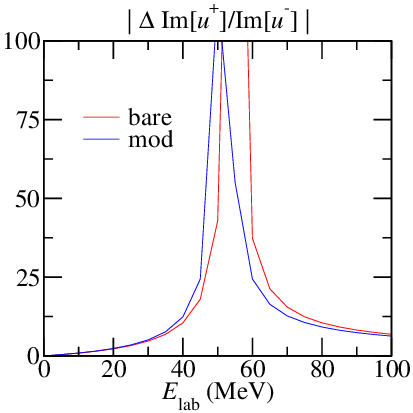}&
\includegraphics[scale=0.7]{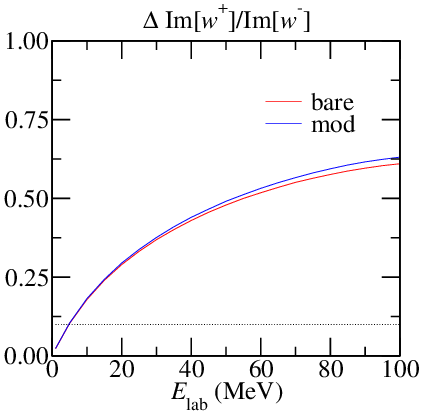}\tabularnewline
\end{tabular}

\caption{The similar analysis as Fig.\,\ref{fig:10-to-5} for the $x^{-}$
and $x^{+}$ amplitudes with the Yukawa mass parameter in $V_{-1,\mathrm{SR}}^{\mathrm{PV}}$
chosen to be $m=m_{\pi}$. The dotted line marks the $10\%$ level
and it is not visible for the central bottom panel.\,\label{fig:SRmpi}}
\end{figure}

One might be temped to think that, as this bad scaling is due to the
huge difference between the Yukawa mass scales: $m_{\mathrm{SR}}=m_{\rho}$
in $V_{1,\mathrm{SR}}^{\mathrm{PV}}$ and $m_{\mathrm{LR}}=m_{\pi}$
in $V_{-1,\mathrm{LR}}^{\mathrm{PV}}$, then, choosing a smaller mass
scale in $V_{1,\mathrm{SR}}^{\mathrm{PV}}$ should make the pionless
EFT work. The most straightforward choice is $m_{\mathrm{SR}}=m_{\pi}$---as
long as the energy being considered is much smaller than $m_{\pi}$---by
which the $w^{-}$ and $w^{\pi}$ amplitudes become identical. In
this case, as Fig.\,\ref{fig:SRmpi} shows (the {}``mod'' calculations
are very close the the {}``bare'' ones since the form factors do
not cutoff the LR interaction too much), the scaling between $x^{+}$
and $x^{-}$ becomes extremely bad, and this completely spoils the
nice reduction scheme of 10-to-5 LECs. For the $v$ and $w$ amplitudes,
the $10\%$ scaling deviation only allows $E_{\mathrm{lab}}\lesssim5\,\mbox{MeV}$;
for the $u$ amplitude, it is even worse (the divergence at $E_{\mathrm{lab}}\sim50\mbox{--}60\,\mbox{MeV}$
is because Im{[}$u^{-}$] crosses over the zero value). The reason
is again that, by making the effective range of $V_{1,\mathrm{SR}}^{\mathrm{PV}}$
longer, the difference between $\langle\bm y_{m+}\rangle$ and $\langle\bm y_{m-}\rangle$
gets enhanced as the long-range part of the wave function, having
a larger gradient, gets a bigger weight. Therefore, the price to pay
for making the pionless EFT work is to keep all the 10 LECs as independent
parameters. Clearly, the more economic choice would be keeping the
OPE part explicitly and having the 10-to-5 reduction work by choosing
$m_{\mathrm{SR}}=m_{\rho}$ for $V_{1,\mathrm{SR}}^{\mathrm{PV}}$.

To summarize the discussions so far, we conclude that at low energy,
i.e., $E_{\mathrm{lab}}\lesssim90\,\mbox{MeV}$ for $^{1}S_{0}$--$^{3}P_{0}$
transitions and $E_{\mathrm{lab}}\lesssim40\,\mbox{MeV}$ for $^{3}S_{1}$--$^{1}P_{1}$
and $^{3}S_{1}$--$^{3}P_{1}$ transitions, an PV potential formulated
in EFT requires at least 6 parameters: 5 LECs plus $\wt{C}_{6}^{\pi}\propto h_{\pi}^{1}$.
One may think that this conclusion makes  $V_{\mathrm{EFT}}^{\mathrm{PV}}$
equivalent to $V_{\mathrm{OME}}^{\mathrm{PV}}$, which also has six
PV parameters: 5 heavy-meson-nucleon couplings (ignoring $h_{\rho}^{1'}$)
plus $h_{\pi}^{1}$. However, this statement is not true in general,
because the six-parameter EFT framework only works under the assumption
of the $S$--$P$ dominance, but the OME framework is not limited
by this requirement. Furthermore, the OME framework implies some prescribed
relationships, Eqs.\,(\ref{eq:w-collapse}, \ref{eq:r-collapse},
\ref{eq:wtoDDH}, and \ref{eq:rtoDDH}), between $C$'s and $\wt{C}$'s
in the EFT framework; unfortunately, these relationships can not be
justified without going to high energy. 

At last, we come to the determination of the Danilov parameters, $\lambda_{s}^{pp,nn,np}$,
$\lambda_{t}$, and $\rho_{t}$, which will serve as the 5 independent
LECs. The relations between Danilov parameters and the zero-energy
$S$--$P$ amplitudes are given by \begin{subequations}\begin{align}
v^{NN'}(0) & \approx-a_{s}^{NN'}\,\mbox{e}^{i\,(\delta_{3P0}^{NN'}(0)+\delta_{1S0}^{NN'}(0))}\,\lambda_{s}^{NN'}\,,\\
u(0) & \approx-a_{t}\,\mbox{e}^{i\,(\delta_{1P1}(0)+\delta_{3S1}(0))}\,\lambda_{t}\,,\\
w_{\mathrm{SR}}(0) & \approx-a_{t}\,\mbox{e}^{i\,(\delta_{3P1}(0)+\delta_{3S1}(0))}\,\rho_{t}\,,\end{align}
\end{subequations} where $a$ denotes the corresponding scattering
length and $\delta(0)$ the zero-energy phase shift (including the
Coulomb contribution)\,\cite{Desplanques:1978mt}. The notation {}``$w_{\mathrm{SR}}$''
means the OPE-subtracted $^{3}S_{1}$--$^{3}P_{1}$ amplitude. Using
the values $a_{s}^{pp}=-7.8064\,\mbox{fm}$, $a_{s}^{nn}=-18.487\,\mbox{fm}$,
$a_{s}^{np}=-23.7318\,\mbox{fm}$, and $a_{t}=5.4192\,\mbox{fm}$,
we get the dimensionless Danilov parameters\begin{subequations}\begin{align}
\mN\lambda_{s}^{pp} & =5.507\times10^{-3}\,(\wt{D}_{v}^{pp}+0.789\, D_{v}^{pp}-1.655\,\wt{C}_{2}^{2\pi})\,,\\
\mN\lambda_{s}^{nn} & =5.796\times10^{-3}\,(\wt{D}_{v}^{nn}+0.792\, D_{v}^{nn}-1.648\,\wt{C}_{2}^{2\pi})\,,\\
\mN\lambda_{s}^{np} & =5.778\times10^{-3}\,(\wt{D}_{v}^{np}+0.809\, D_{v}^{np})\,,\\
\mN\lambda_{t} & =-1.462\times10^{-3}\,(\wt{D}_{u}-2.230\, D_{u})\,,\\
\mN\rho_{t} & =3.108\times10^{-3}\,(\wt{D}_{w}+0.604\, D_{w}-1.771\,\wt{C}_{6}^{2\pi})\,;\end{align}
\end{subequations} for the {}``bare'' case; and \begin{subequations}\begin{align}
\mN\bar{\lambda}_{s}^{pp} & =3.268\times10^{-3}\,(\wt{D}_{v}^{pp}+0.849\, D_{v}^{pp}-1.260\,\wt{C}_{2}^{2\pi})\,,\\
\mN\bar{\lambda}_{s}^{nn} & =3.809\times10^{-3}\,(\wt{D}_{v}^{nn}+0.853\, D_{v}^{nn}-1.237\,\wt{C}_{2}^{2\pi})\,,\\
\mN\bar{\lambda}_{s}^{np} & =3.772\times10^{-3}\,(\wt{D}_{v}^{np}+0.871\, D_{v}^{np})\,,\\
\mN\bar{\lambda}_{t} & =-0.867\times10^{-3}\,(\wt{D}_{u}-2.425\, D_{u})\,,\\
\mN\bar{\rho}_{t} & =2.003\times10^{-3}\,(\wt{D}_{w}+0.664\, D_{w}-1.586\,\wt{C}_{6}^{2\pi})\,,\end{align}
\end{subequations}for the {}``mod'' case.

In these expressions, one sees that the scaling factors between $D$'s
and $\wt{D}$'s are not $1$ as anticipated by the ZRA with the absence
of hard core. The most striking case, which amounts to $2.230$ (or
$2.425$ for the {}``mod'' case), happens for the $^{1}P_{1}$--$^{3}S_{1}$
transition. This in fact shows the attractiveness of the hybrid EFT
treatment for nuclear PV: one gets the strong dynamics right without
having to go to a higher order in $Q$, where the proliferation of
needed PV parameters is totally undesirable.

\section{Parity-Violating Observables in Two-Nucleon Systems~\label{sec:observables}}

Having determined a minimal set of PV parameters, \textit{\emph{i.e.}},
$\mN\lambda_{s}^{pp,nn,np}$, $\mN\lambda_{t}$, $\mN\rho_{t}$ and
$\wt{C}_{6}^{\pi}$, to describe the low-energy PV phenomena, we will
use them in this section to express the PV observables which have
been or will be measured in two-body systems. As analyses of these
observables have been quite extensively discussed in the $V_{\textrm{OME}}^{\textrm{PV}}$
framework, we refer most of the details which also apply to the EFT
framework to literature and only highlight the new results and the
comparison with the old framework.

\subsection{$A_{L}^{\vec{p}p}$ in $\vec{p}\, p$ scattering}

The {}``nuclear'' total asymmetry for $\vec{p}\, p$ scattering~\cite{Simonius:1973tj,Simonius:1974,Simonius:1988rg,Nessi-Tedaldi:1988ak,Driscoll:1989hg,Carlson:2001ma}, 

\begin{equation}
A_{L}^{\vec{p}p}(E)=\frac{\textrm{Im}\left[\wt{\mathcal{M}}_{10,00}(E,0)+\wt{\mathcal{M}}_{00,10}(E,0)\right]}{\textrm{Im}\left[{\displaystyle {\sum_{S,M_{S}}}}\mathcal{M}_{SM_{S},SM_{S}}(E,0)\right]}\,,\end{equation}
is defined through the Coulomb-subtracted, forward ($\theta=0$) scattering
amplitude $\mathcal{M}(E,0)$, where the subscript pair $S'M'_{S},SM_{S}$
denotes the final and initial two-body spin states, respectively.
As the Coulomb scattering amplitude, $M^{\textrm{C}}$, diverges at
the forward angle, the total asymmetry (with full $4\,\pi$ angular
coverage) can only be well-defined after this singular piece is subtracted:
$\mathcal{M}\equiv M-M^{\textrm{C}}$. One should note that $A_{L}^{\vec{p}p}$
is not a quantity an experiment directly measures; a theoretical correction
is needed to fold an experimental result into $A_{L}^{\vec{p}p}(E)$
(see, \textit{\emph{e.g.}}, Refs.~\cite{Driscoll:1989hg,Carlson:2001ma}
for more discussions).

Currently, there are two low-energy data points at $13.6\,\textrm{MeV}$
and $45\,\textrm{MeV}$ which give $A_{L}^{\vec{p}p}=-(0.93\pm0.21)\times10^{-7}$~\cite{Eversheim:1991tg,Haeberli:1995}
and $-(1.57\pm0.23)\times10^{-7}$~\cite{Kistryn:1987tq}, respectively.
These supersede the earlier, less accurate experiments at $15$ and
$45\,\textrm{MeV}$ which yield $-(1.7\pm0.8)\times10^{-7}$~\cite{Potter:1974gu,Nagle:1979}
and $-(2.31\pm0.89)\times10^{-7}$~\cite{Balzer:1980dn,Balzer:1985au},
respectively. At higher energy, there is one measurement at $221\,\textrm{MeV}$,
yielding $+(0.84\pm0.29)\times10^{-7}$~\cite{Berdoz:2001nu,Berdoz:2002sn};
it is motivated by the theoretical prediction that this would uniquely
determine the PV $\rho$-exchange coupling constant $h_{\rho}^{pp}\equiv h_{\rho}^{0}+h_{\rho}^{1}+h_{\rho}^{2}/\sqrt{6}$
in $V_{\textrm{OME}}^{\textrm{PV}}$~\cite{Simonius:1988rg}. 

\begin{table}

\caption{Analysis of $A_{L}^{\vec{p}p}$ decomposed in partial waves. Each
entry denotes the multiplicative coefficient for the corresponding
PV coupling constant. The full result is the sum of every {}``entry$\times$coupling''
in the same row. The last column {}``OME-m'' shows the numerical
value of $A_{L}^{\vec{p}p}$ by performing the OME-mapping to the
EFT result with the PC and PV parameters specified in Tabs.\,\ref{cap:strong-parameters}
and \ref{cap:weak-parameters}.~\label{cap:pp_AL}}

\begin{longtable}{|c|c||c|c|c|c|c|c|c|c|c||r|}
\hline 
\multicolumn{1}{|c}{}&
\multicolumn{1}{c||}{}&
\multicolumn{3}{c|}{$^{1}S_{0}\textrm{--}^{3}P_{0}\,(\times10^{-3})$}&
\multicolumn{3}{c|}{$^{1}D_{2}\textrm{--}^{3}P_{2}\,(\times10^{-3})$}&
\multicolumn{3}{c||}{$^{1}D_{2}\textrm{--}^{3}F_{2}\,(\times10^{-3})$}&
OME-m\tabularnewline
\cline{3-5} \cline{6-8} \cline{9-11} 
\multicolumn{1}{|c}{}&
\multicolumn{1}{c||}{}&
$D_{v}^{pp}$&
$\wt{D}_{v}^{pp}$&
$\wt{C}_{2}^{2\pi}$&
$D_{v}^{pp}$&
$\wt{D}_{v}^{pp}$&
$\wt{C}_{2}^{2\pi}$&
$D_{v}^{pp}$&
$\wt{D}_{v}^{pp}$&
$\wt{C}_{2}^{2\pi}$&
$(\times10^{-7})$\tabularnewline
\hline
\hline 
\multicolumn{1}{|c|}{\multirow{2}{0.75cm}{$13.6$}}&
bare&
$-1.980$&
$-2.476$&
$4.010$&
$-0.005$&
$0.006$&
$-0.012$&
$0.001$&
$-0.001$&
$0.002$&
$-0.971$\tabularnewline
\cline{2-2} \cline{3-3} \cline{4-4} \cline{5-5} \cline{6-6} \cline{7-7} \cline{8-8} \cline{9-9} \cline{10-10} \cline{11-11} \cline{12-12} 
\multicolumn{1}{|c|}{}&
mod&
$-1.398$&
$-1.617$&
$1.885$&
$-0.004$&
$0.004$&
$-0.010$&
$0.000$&
$-0.001$&
$0.001$&
$-0.960$\tabularnewline
\cline{1-1} 
\cline{2-2} \cline{3-3} \cline{4-4} \cline{5-5} \cline{6-6} \cline{7-7} \cline{8-8} \cline{9-9} \cline{10-10} \cline{11-11} \cline{12-12} 
\multicolumn{1}{|c|}{\multirow{2}{0.75cm}{$45$}}&
bare&
$-3.686$&
$-4.476$&
$7.026$&
$-0.122$&
$0.132$&
$-0.243$&
$0.027$&
$-0.033$&
$0.064$&
$-1.746$\tabularnewline
\cline{2-2} \cline{3-3} \cline{4-4} \cline{5-5} \cline{6-6} \cline{7-7} \cline{8-8} \cline{9-9} \cline{10-10} \cline{11-11} \cline{12-12} 
\cline{2-2} \cline{3-3} \cline{4-4} \cline{5-5} \cline{6-6} \cline{7-7} \cline{8-8} \cline{9-9} \cline{10-10} \cline{11-11} \cline{12-12} 
\multicolumn{1}{|c|}{}&
mod&
$-2.582$&
$-2.868$&
$2.842$&
$-0.089$&
$0.094$&
$-0.180$&
$0.019$&
$-0.023$&
$0.050$&
$-1.662$\tabularnewline
\cline{1-1} 
\cline{2-2} \cline{3-3} \cline{4-4} \cline{5-5} \cline{6-6} \cline{7-7} \cline{8-8} \cline{9-9} \cline{10-10} \cline{11-11} \cline{12-12} 
\multicolumn{1}{|c|}{\multirow{2}{0.75cm}{$221$}}&
bare&
$-0.069$&
$-0.073$&
$0.112$&
$-2.749$&
$2.618$&
$-3.784$&
$0.164$&
$-0.388$&
$0.633$&
$0.426$\tabularnewline
\cline{2-2} \cline{3-3} \cline{4-4} \cline{5-5} \cline{6-6} \cline{7-7} \cline{8-8} \cline{9-9} \cline{10-10} \cline{11-11} \cline{12-12} 
\cline{2-2} \cline{3-3} \cline{4-4} \cline{5-5} \cline{6-6} \cline{7-7} \cline{8-8} \cline{9-9} \cline{10-10} \cline{11-11} \cline{12-12} 
\multicolumn{1}{|c|}{}&
mod&
$-0.046$&
$-0.043$&
$0.015$&
$-1.888$&
$1.678$&
$-1.636$&
$0.086$&
$-0.270$&
$0.420$&
$0.853$\tabularnewline
\hline
\end{longtable}
\end{table}

The EFT analysis of $A_{L}^{\vec{p}p}$ for the low-energy data points
is tabulated in Tab.~\ref{cap:pp_AL}. Apparently, the observables
are dominated by the $S$--$P$ transition. Using the Danilov parameters
obtained in the last section, we find that\begin{align}
A_{L}^{\vec{p}p}(13.6\,\mbox{MeV}) & =-0.449\,\mN\lambda_{s}^{pp}+(-0.035\, D_{v}^{pp}-0.088\,\wt{C}_{2}^{2\pi})\times10^{-3}\quad(\mbox{bare}),\nonumber \\
 & \mbox{or}-0.445\,\mN\bar{\lambda}_{s}^{pp}+(-0.032\, D_{v}^{pp}-0.157\,\wt{C}_{2}^{2\pi})\times10^{-3}\quad(\mbox{mod});\\
A_{L}^{\vec{p}p}(45\,\mbox{MeV}) & =-0.795\,\mN\lambda_{s}^{pp}+(-0.329\, D_{v}^{pp}-0.395\,\wt{C}_{2}^{2\pi})\times10^{-3}\quad(\mbox{bare}),\nonumber \\
 & \mbox{or}-0.771\,\mN\bar{\lambda}_{s}^{pp}+(-0.276\, D_{v}^{pp}-0.813\,\wt{C}_{2}^{2\pi})\times10^{-3}\quad(\mbox{mod}).\end{align}
The correction terms, enclosed in parentheses, are in general quite
small except for the TPE parts in the {}``mod'' calculations, so
we conclude that these two data only depend on one single parameter,
$\mN\lambda_{s}^{pp}$. In fact, the theoretical prediction for the
ratio of $A_{L}^{\vec{p}p}(45\,\mbox{MeV})/A_{L}^{\vec{p}p}(13.6\,\mbox{MeV})\approx-0.795/-0.449=1.77$
(or $-0.771/-0.445=1.73$ for the {}``mod'' case) agrees very well
with the experimental result $\approx-1.57/-0.93=1.69$ (discarding
errors). Furthermore, one can see from the comparison between the
{}``bare'' and {}``mod'' results that even though $\mN\lambda_{s}^{pp}$
and $\mN\bar{\lambda}_{s}^{pp}$ are defined in different models,
the expressions for $A_{L}^{\vec{p}p}$ in terms of them are almost
model-independent. This justifies the advantage of using the Danilov
parameters instead of the LECs $C$'s, $\wt{C}'s$, etc.

Even though it is doubtful that the $V_{\mathrm{EFT}}^{\textrm{PV}}$
of order $O(Q)$ is sufficient for analyzing the high-energy datum
at $221\,\textrm{MeV}$, it is nonetheless interesting to see how
the analysis turns out to be. The result, also shown in Tab.~\ref{cap:pp_AL},
indicates a quite different feature. The $D\textrm{--}P$ amplitude
becomes the most dominant contribution, with the $D$--$F$ one also
being non-negligible. Both amplitudes have their own scaling factors
between $D_{v}^{pp}$, $\wt{D}_{v}^{pp}$, and $\wt{C}_{2}^{2\pi}$
components different from the $S$--$P$ amplitude. While it is not
clear if the $D$--$P$ and $D$--$F$ amplitudes can be completely
specified by $D_{v}^{pp}$, $\wt{D}_{v}^{pp}$, and $\wt{C}_{2}^{2\pi}$,
they certainly can not be uniquely fixed by the only high-energy measurement.

By OME-mapping the EFT results in Tab.~\ref{cap:pp_AL}, our calculations
are checked with literature. The {}``bare+DDH-best'' results are
consistent with works such as Refs.~\cite{Simonius:1973tj,Simonius:1974,Simonius:1988rg,Nessi-Tedaldi:1988ak},
given that different strong potential models are used. The {}``mod+DDH-adj.''
results agree well with Refs.~\cite{Driscoll:1989hg,Carlson:2001ma};
the small difference is because we do not use a different mass and
a cutoff factor for the $\omega$ meson. 

It is worth to point out that the analysis by Carlson et al\emph{.}\,\cite{Carlson:2001ma},
which is based on a OME framework with two independent parameters
$h_{\rho}^{pp}$ and $h_{\omega}^{pp}$, claims a good fit to both
low- and high-energy data. Unfortunately, due to the lack of more
high energy data, it is currently impossible to verify this fit along
with its important dynamical assumptions---the monopole form factors
and big isovector tensor coupling $\chi_{\rho}$---within the EFT
framework. On the other hand, their fitted PV $\omega NN$ coupling
constant, $h_{\omega}^{pp}=h_{\omega}^{0}+h_{\omega}^{1}$, is only
marginally consistent with most hadronic perditions and needs further
clarification. For these issues, we refer to Ref.~\cite{Liu:2005sn}
for more discussions.

Finally, we turn our attention to the TPE contributions, which have
not received extensive study and are left out while we do the OME-mapping
in Tab.~\ref{cap:pp_AL}. By writing out $\wt{C}_{2}^{2\pi}$ in
term of $h_{\pi}^{1}$ explicitly and assuming the DDH best value
for $h_{\pi}^{1}$, one sees the asymmetry in the {}``bare'' case
increased by $\sim70\%$ for the $13.6$ and $45\,\mbox{MeV}$ data
points and $\sim60\%$ for the $221\,\mbox{MeV}$ one. In the {}``mod''
case, the increases are $\sim30\%$ and $\sim20\%$ for low- and high-energy
cases, respectively. Although these corrections seem quite big, one
should remember they are not the full TPE corrections, as part of
the TPE contributions are buried in the SR interaction. But since
$V_{1,\mathrm{MR}}^{\mathrm{PV}}$ and $V_{1,\mathrm{SR}}^{\mathrm{PV}}$
have the same power counting, it is not unnatural to expect individual
terms of similar magnitude. Another remark concerns the general trends
that the TPE enhances the asymmetry and it is the the low-energy cases
that get more boost than the high-energy ones. These points have also
been noticed in Ref.\,\cite{Liu:2005sn}, where part of the TPE contribution
is accounted for by formulating it as a $\rho$-resonance.

\subsection{$\phi_{n}^{\vec{n}p}$ and $P_{n}^{np}$ in neutron transmission
through para-hydrogen}

It was first pointed by Michel~\cite{Michel:1964} and later on by
Stodolsky~\cite{Stodolsky:1974hm,Stodolsky:1981vn} that nuclear
parity violation can be studied through low-energy neutron transmission,
where the whole process acts like optics. The observables could be
a spin rotation, $\phi_{n}$, about the longitudinal axis (assumed
to be $\hat{z}$) for the transversely-polarized neutron, or a net
longitudinal polarization, $P_{n}$, that an unpolarized neutron beam
picks up when traversing through medium---the latter is equivalent
to the asymmetry in cross section for the longitudinally-polarized
neutron scattering, $A_{L}^{\vec{n}p}$. These quantities per unit
length (assuming the target is uniform), $d\,\phi_{n}/d\, z$ and
$d\, P_{n}/d\, z$, can be related to the PV forward scattering amplitude
$\wt{M}(E,0)$ by \begin{align}
\frac{d\,\phi_{n}}{d\, z} & =-\frac{2\,\pi}{k}\, N\,\textrm{Re}(\wt{M}_{+z}(E,0)-\wt{M}_{-z}(E,0))\,,\label{eq:phiPV}\\
\frac{d\, P_{n}}{d\, z} & =-\frac{2\,\pi}{k}\, N\,\textrm{Im}(\wt{M}_{+z}(E,0)-\wt{M}_{-z}(E,0))\,,\label{eq:PPV}\end{align}
where $k$ is the magnitude of the neutron momentum, $N$ is the target
number density, and the subscript $\pm z$ denotes the direction of
neutron polarization.

For a thermal neutron beam, $E_{n}\approx0.025\,\textrm{eV}$, transmitting
through liquid para-hydrogen, $N=0.24\times10^{23}/\textrm{cm}^{3}$,
the EFT analysis of $d\,\phi_{n}/d\, z$ and $d\, P_{n}/d\, z$ is
tabulated in Tab.~\ref{cap:np_transmission}. At thermal energy,
the magnitude of $d\, P_{n}^{np}/d\, z$ is about four orders of magnitude
smaller than $d\,\phi_{n}^{\vec{n}p}/d\, z$. When neutron energy
is further decreased, $d\,\phi_{n}^{\vec{n}p}/d\, z$ stays constant,
but $d\, P_{n}^{np}/d\, z$ drops as $\sqrt{E_{n}}$; therefore, the
spin-rotation measurement is more feasible for low-energy neutrons.
This trend is consistent with the argument made by Stodolsky~\cite{Stodolsky:1981vn}
about the elastic scattering. It is also pointed out in Ref.~\cite{Stodolsky:1981vn}
that exothermic processes, \emph{}i.e., inelastic exit channels, can
possibly result in a non-vanishing $d\, P_{n}/d\, z$ at zero energy.
However, it is not the case for $n\, p$ scattering, where the only
exothermic reaction, $n\, p\rightarrow d\,\gamma$ (will be discussed
in Sec.~\ref{sub:A_g}), does not lead to a total asymmetry, as remarked
in Ref.\,\cite{Liu:2006qp}.

\begin{table}

\caption{Analysis of (I) $d\,\phi_{n}^{\vec{n}p}(\textrm{th.})/d\, z$ in
rad/m and (II) $d\, P_{n}^{np}(\textrm{th.})/d\, z$ in $10^{-4}/\textrm{m}$
decomposed in partial waves. See Tab.\, \ref{cap:pp_AL} for the
explanation of tabularization.~\label{cap:np_transmission}}

\begin{tabular}{|c|c||d|d|d|d|d|d|d|d||d|}
\hline 
\multicolumn{1}{|c}{}&
\multicolumn{1}{c||}{}&
\multicolumn{2}{c|}{$^{1}S_{0}\textrm{--}^{3}P_{0}\,(\times10^{-2})$}&
\multicolumn{2}{c|}{$^{3}S_{1}\textrm{--}^{1}P_{1}\,(\times10^{-2})$}&
\multicolumn{4}{c||}{$^{3}S_{1}\textrm{--}^{3}P_{1}\,(\times10^{-2})$}&
\multicolumn{1}{c|}{OME-m}\tabularnewline
\cline{3-4} \cline{5-6} \cline{7-10} 
\cline{3-3} \cline{4-4} \cline{5-5} \cline{6-6} \cline{7-7} \cline{8-8} \cline{9-9} 
\multicolumn{1}{|c}{}&
\multicolumn{1}{c||}{}&
\multicolumn{1}{c|}{$D_{v}^{np}$}&
\multicolumn{1}{c|}{$\wt{D}_{v}^{np}$}&
\multicolumn{1}{c|}{$D_{u}$}&
\multicolumn{1}{c|}{$\wt{D}_{u}$}&
\multicolumn{1}{c|}{$D_{w}$}&
\multicolumn{1}{c|}{$\wt{D}_{w}$}&
\multicolumn{1}{c|}{$\wt{C}_{6}^{\pi}$}&
\multicolumn{1}{c||}{$\wt{C}_{6}^{2\pi}$}&
\multicolumn{1}{c|}{$(\times10^{-7})$}\tabularnewline
\hline
\hline 
\multicolumn{1}{|c|}{\multirow{3}{0.5cm}{(I) }}&
bare&
1.169&
1.444&
-0.186&
0.083&
0.214&
0.355&
0.286&
-0.628&
6.711\tabularnewline
\cline{2-2} \cline{3-3} \cline{4-4} \cline{5-5} \cline{6-6} \cline{7-7} \cline{8-8} \cline{9-9} \cline{10-10} \cline{11-11} 
\cline{2-2} \cline{3-3} \cline{4-4} \cline{5-5} \cline{6-6} \cline{7-7} \cline{8-8} \cline{9-9} \cline{10-10} \cline{11-11} 
\multicolumn{1}{|c|}{}&
mod&
0.821&
0.943&
-0.120&
0.049&
0.152&
0.229&
0.284&
-0.363&
5.149\tabularnewline
\cline{1-1} 
\cline{2-2} \cline{3-3} \cline{4-4} \cline{5-5} \cline{6-6} \cline{7-7} \cline{8-8} \cline{9-9} \cline{10-10} \cline{11-11} 
\multicolumn{1}{|c|}{\multirow{3}{0.5cm}{(II) }}&
bare&
4.805&
5.939&
0.175&
-0.079&
-0.202&
-0.334&
-0.269&
0.591&
2.165\tabularnewline
\cline{2-2} \cline{3-3} \cline{4-4} \cline{5-5} \cline{6-6} \cline{7-7} \cline{8-8} \cline{9-9} \cline{10-10} \cline{11-11} 
\cline{2-2} \cline{3-3} \cline{4-4} \cline{5-5} \cline{6-6} \cline{7-7} \cline{8-8} \cline{9-9} \cline{10-10} \cline{11-11} 
\multicolumn{1}{|c|}{}&
mod&
3.378&
3.878&
0.113&
-0.047&
-0.143&
-0.215&
-0.267&
0.341&
-2.220\tabularnewline
\hline
\end{tabular}
\end{table}

In this case, all three different $S\textrm{--}P$ amplitudes come
into play, and the results in terms of Danilov parameters and $\wt{C}_{6}^{\pi}$
are \begin{align}
\frac{d\,\phi_{n}^{\vec{n}p}(\mathrm{th.})}{d\, z}\Big|_{\mathrm{m/rad}} & =0.286\,\wt{C}_{6}^{\pi}+2.500\,\mN\lambda_{s}^{np}-0.571\,\mN\lambda_{t}+1.412\,\mN\rho_{t}+(0.000)\quad(\mbox{bare});\nonumber \\
 & \mbox{or }0.284\,\wt{C}_{6}^{\pi}+2.500\,\mN\bar{\lambda}_{s}^{np}-0.571\,\mN\bar{\lambda}_{t}+1.412\,\mN\bar{\rho}_{t}+(0.000)\quad(\mbox{mod}).\end{align}
Because this process is close to zero energy, it is not a surprise
that the Danilov parameters work extremely well (almost no error).
Also, this expression is model-independent, no matter it is for the
{}``bare'' or the {}``mod'' calculation.

By OME-mapping the EFT results, the {}``bare+DDH-best'' value, $d\,\phi_{n}^{\vec{n}p}(\textrm{th.})/d\, z\simeq6.71\times10^{-7}\,\textrm{rad}/\textrm{m}$,
is about $20\%$ smaller in magnitude than an early prediction using
the Paris potential~\cite{Avishai:1984mu}, and with a different
sign. Thus, we confirm the assertion of Ref.~\cite{Schiavilla:2004wn}
about the sign problem in Ref.~\cite{Avishai:1984mu}. As for the
{}``mod+DDH-adj'' value, $\simeq5.15\times10^{-7}\,\textrm{rad}/\textrm{m}$,
it agrees well with Ref.~\cite{Schiavilla:2004wn}. If these numerical
estimates are not too far off, the plan of doing such an experiment
aiming at a $2.7\times10^{-7}\,\textrm{rad/m}$ precision~\cite{Markov:2002}
at the Spallation Neutron Source (SNS) will certainly provide a valuable
data point.

The TPE contribution enters through the $^{3}S_{1}$--$^{3}P_{1}$
transition. Because $\wt{C}_{6}^{\pi}$ and $\wt{C}_{6}^{2\pi}$ have
the same sign, Tab.\,\ref{cap:np_transmission} suggests that the
TPE reduces the OPE contribution which dominates the above OME-m estimates.
The correction is about $-15\%$ for the {}``bare'' calculation,
and $-10\%$ for the {}``mod'' calculation. This $\sim10\%$ correction
is consistent with the qualitative power-counting argument that the
TPE contribution is smaller than the leading OPE one by an order of
magnitude.

\subsection{$P_{\gamma}^{np}$ in $n\, p\rightarrow d\,\gamma$ and $A_{L}^{\vec{\gamma}d}$
in $\vec{\gamma}\, d\rightarrow n\, p$\,\label{sub:P_g}}

Low-energy radiative neutron capture mainly involves the lowest-order
electromagnetic transitions. For $n\, p\rightarrow d\,\gamma$, it
is $M1$ for the PC part, and $E1$ for the PV part. Since the total
cross section is dominated by the $^{1}S_{0}$-wave scattering, the
non-zero circular polarization takes an approximate simple form as\begin{equation}
P_{\gamma}^{np}=2\,\frac{\bra{\mathcal{D}}|E_{1}|\wt{\ket{^{3}P_{0}}}+\vspace{0.5cm}_{\mathcal{D}}\wt{\bra{^{1}P_{1}}}|E_{1}|\ket{^{1}S_{0}}}{\bra{\mathcal{D}}|M_{1}|\ket{^{1}S_{0}}}\,,\label{eq:P_gamma}\end{equation}
where the double bar {}``$||$'' denotes the reduced matrix element.
In this case, the observable depends on the $^{1}S_{0}\textrm{--}^{3}P_{0}$
and the deuteron $\mathcal{D}\textrm{--}^{1}P_{1}$ admixtures. It
is important to recognize that we rely on the Siegert theorem~\cite{Siegert:1937yt},
through which the $E_{1}$ operator is related to the charge dipole
operator $C_{1}$, to calculate the $E1$ matrix elements. This manipulation
not only implicitly includes most $\mathcal{O}(v/c)$ meson exchange
currents, but also imposes a $\Delta S=0$ spin selection rule as
shown in Eq.\,(\ref{eq:P_gamma}). 

\begin{table}

\caption{Analysis of $P_{\gamma}^{np}(\textrm{th.})$ decomposed in partial
waves. See Tab.\, \ref{cap:pp_AL} for the explanation of tabularization.~\label{cap:np_Pg}}

\begin{tabular}{|c|c|c|c|c||c|}
\hline 
&
\multicolumn{2}{c|}{$^{1}S_{0}\textrm{--}^{3}P_{0}$ mix. $(\times10^{-3})$}&
\multicolumn{2}{c||}{$\mathcal{D}\textrm{--}^{1}P_{1}$ mix. $(\times10^{-3})$}&
\multicolumn{1}{c|}{OME-m}\tabularnewline
\cline{2-3} \cline{4-5} 
&
$D_{v}^{np}$&
$\wt{D}_{v}^{np}$&
$D_{u}$&
$\wt{D}_{u}$&
\multicolumn{1}{c|}{$(\times10^{-7})$}\tabularnewline
\hline
\hline 
bare&
$-0.751$&
$-0.935$&
$2.166$&
$-0.980$&
$0.247$\tabularnewline
\hline 
mono&
$-0.525$&
$-0.607$&
$1.391$&
$-0.580$&
$0.520$\tabularnewline
\hline
\end{tabular}
\end{table}

For thermal neutron, the EFT analysis is tabulated in Tab.~\ref{cap:np_Pg}.
Although the observable $P_{\gamma}^{np}$ is not determined directly
by the scattering amplitudes---instead, by the parity admixtures which
do not have a trivial relation to the scattering amplitudes in general---the
Danilov parameters still do a good job \begin{align}
P_{\gamma}^{np}(\mbox{th.}) & =-0.161\,\mN\lambda_{s}^{np}+0.670\,\mN\lambda_{t}+(0.005\, D_{v}^{np}+0.019\, D_{u})\times10^{-3}\quad(\mbox{bare});\nonumber \\
 & \mbox{or }-0.161\,\mN\bar{\lambda}_{s}^{np}+0.669\,\mN\bar{\lambda}_{t}+(0.004\, D_{v}^{np}+0.016\, D_{u})\times10^{-3}\quad(\mbox{mod}).\end{align}
The reason is mainly because the process occurs at a very low energy
and the deuteron is a loosely bound state; both are not far from the
zero-energy limit. 

The OME-mapping gives the {}``bare+DDH-best'' value $P_{\gamma}^{np}(\textrm{th.})=2.5\times10^{-8}$
which agrees well with a recent calculation~\cite{Hyun:2004xp} and
is also consistent with pre-80's predictions, e.g., Refs.~\cite{Tadic:1968,Danilov:1971fh,Lassey:1975,Desplanques:1975,Craver:1976am},
around $(2\textrm{--}5)\times10^{-8}$, (see Ref.~\cite{Lassey:1976kv}
for a summary). For the {}``mod+DDH-adj.'' value $P_{\gamma}^{np}(\textrm{th.})=0.52\times10^{-7}$,
we have an excellent check with Ref.~\cite{Schiavilla:2004wn}.

Historically, the first measurement of $P_{\gamma}^{np}(\textrm{th.})$
done by the Leningrad group reports a result $-(1.3\pm0.45)\times10^{-6}$~\cite{Lobashov:1972},
which not only exceeds most theoretical predictions by two orders
of magnitude but also has an opposite sign. The follow-up experiment
does correct the sign problem; however, the published result $P_{\gamma}^{np}(\textrm{th.})=(1.8\pm1.8)\times10^{-7}$~\cite{Knyaz'kov:1984}
still has too large an error bar. In order to circumvent the difficulty
of measuring a circular polarization, the inverse process, the asymmetry
$A_{L}^{\vec{\gamma}d}$ in deuteron photo-disintegration, $\vec{\gamma}\, d\rightarrow n\, p$,
can be a good alternative. By detailed balancing, $A_{L}^{\vec{\gamma}d}=P_{\gamma}^{np}$
if all kinematics are exactly reversed. One can show that, for photon
with energy of $1.32\,\textrm{keV}$ above the threshold, $A_{\gamma}^{\vec{\gamma}d}(1.32\,\textrm{keV}+)=P_{\gamma}^{np}(\textrm{th.})$.

As demonstrated in several theoretical works~\cite{Lee:1978kh,Oka:1983sp,Fujiwara:2004zg,Liu:2004zm,Schiavilla:2004wn},
the asymmetry $A_{L}^{\vec{\gamma}d}$ gets larger when approaching
the threshold, but on the other hand, the total cross section gets
smaller. There are two data points reported in 80s: $(2.7\pm2.8)\times10^{-6}$
at $E_{\gamma}=4.1\,\textrm{MeV}$ and $(7.7\pm5.3)\times10^{-6}$
at $E_{\gamma}=3.2\,\textrm{MeV}$~\cite{Earle:1988fc,Alberi:1988fd};
though they qualitatively justify the statement above, the precisions
are too low to be of use. With various groups showing interests of
new measurements (e.g., Ref.~\cite{jlab-lett00}), it is important
to decide the best photon energy (should not be too far from the threshold
for larger asymmetry) to be employed. 

An important point to note for this particular observable is that,
unlike the case for neutron transmission, the $\mathcal{D}\textrm{--}^{1}P_{1}$
admixture has an important contribution so that the model dependence
is worrisome. The situation is most clear when comparing with other
semi- and non-local potential model calculations. As shown in Refs.~\cite{Schiavilla:2004wn,Hyun:2004xp},
the CD-Bonn and Bonn-B calculations give predictions two times larger,
and the Bonn calculation even enhances by an order of magnitude. The
difference is mostly due to the large variations of these models in
the $^{1}P_{1}$ channel at short distances~\cite{Hyun:2004xp}.
In this sense, a well-determined $\lambda_{t}$, if ever possible,
can be used in a reversed way to constrain strong potential models.

\subsection{$A_{\gamma}$ in $\vec{n}\, p\rightarrow d\,\gamma$~\label{sub:A_g}}

By the same approximation as in the above subsection, the photon asymmetry
in $\vec{n}\, p\rightarrow d\,\gamma$, $A_{\gamma}^{\vec{n}p}$,
which is defined through $d\,\sigma_{\pm}(\theta)/d\,\Omega\propto1\pm A_{\gamma}^{\vec{n}p}\,\cos\theta$,~%
\footnote{From this expression, one clearly sees $\int\, d\,\Omega\, d\,\sigma_{+}(\theta)=\int\, d\,\Omega\, d\,\sigma_{-}(\theta)$.
This confirms the earlier statement that $A_{L}^{\vec{n}p}$ vanishes
at zero energy, even if an exothermic process exists. %
} can be expressed as\begin{equation}
A_{\gamma}^{\vec{n}p}=-\sqrt{2}\,\frac{\bra{\mathcal{D}}|E_{1}|\wt{\ket{^{3}P_{1}}}+\vspace{0.5cm}_{\mathcal{D}}\wt{\bra{^{3}P_{1}}}|E_{1}|\ket{^{3}S_{1}}}{\bra{\mathcal{D}}|M_{1}|\ket{^{1}S_{0}}}\,.\end{equation}
Unlike the case for $P_{\gamma}^{np}$, it is the $^{3}S_{1}\textrm{--}^{3}P_{1}$
and $\mathcal{D}\textrm{--}^{3}P_{1}$ admixtures that contribute
in this case.

\begin{table}

\caption{Analysis of $A_{\gamma}^{\vec{n}p}(\textrm{th.})$ decomposed in
partial waves. See Tab.\, \ref{cap:pp_AL} for the explanation of
tabularization.~\label{cap:np_Ag}}

\begin{tabular}{|c|c|c|c|c|c|c|c|c||c|}
\hline 
&
\multicolumn{4}{c|}{$^{3}S_{1}\textrm{--}^{3}P_{1}$ mix. $(\times10^{-3})$}&
\multicolumn{4}{c||}{$\mathcal{D}\textrm{--}^{3}P_{1}$ mix. $(\times10^{-3})$}&
OME-m\tabularnewline
\cline{2-5} \cline{6-9} 
&
$D_{w}$&
$\wt{D}_{w}$&
$\wt{C}_{6}^{\pi}$&
$\wt{C}_{6}^{2\pi}$&
$D_{w}$&
$\wt{D}_{w}$&
$\wt{C}_{6}^{\pi}$&
$\wt{C}_{6}^{2\pi}$&
$(\times10^{-7})$\tabularnewline
\hline
\hline 
bare&
$-0.108$&
$-0.185$&
$-0.133$&
$0.321$&
$-0.066$&
$-0.103$&
$-0.139$&
$0.193$&
$-0.506$\tabularnewline
\hline 
mod&
$-0.076$&
$-0.118$&
$-0.132$&
$0.172$&
$-0.048$&
$-0.069$&
$-0.138$&
$0.128$&
$-0.486$\tabularnewline
\hline
\end{tabular}
\end{table}

For thermal neutron, the EFT analysis is tabulated in Tab.~\ref{cap:np_Ag}.
Again, the Danilov parameter $\mN\rho_{t}$ pretty much summarizes
the short-distance physics \begin{align}
A_{\gamma}^{\vec{n}p}(\textrm{th.}) & =-0.272\,\wt{C}_{6}^{\pi}-0.093\,\mN\rho_{t}+(0.003\,\wt{C}_{6}^{2\pi})\times10^{-3}\quad(\mbox{bare});\nonumber \\
 & \mbox{or }-0.270\,\wt{C}_{6}^{\pi}-0.093\,\mN\bar{\rho}_{t}+(0.004\,\wt{C}_{6}^{2\pi})\times10^{-3}\quad(\mbox{mod}).\end{align}

The OME-mapping values, $A_{\gamma}^{\vec{n}p}(\textrm{th.})=-5.06\times10^{-8}$
for the {}``bare+DDH-best'' case and $-4.85\times10^{-8}$ for the
{}``bare+DDH-adj.'' case, are consistent with existing predictions,
e.g., Refs.~\cite{Tadic:1968,Danilov:1971fh,Desplanques:1975,Lassey:1976kv,Kaplan:1998xi,Savage:2000iv,Desplanques:2000ej,Hyun:2001yg,Hyun:2004xp}
for the former and Refs.~\cite{Schiavilla:2002uc,Schiavilla:2004wn}
for the latter, respectively. 

The TPE contributions change the above OME-m results somewhat. Their
corrections to the OPE contributions are $-13\%$ and $-8\%$ for
the {}``bare'' and {}``mod'' cases, respectively. This is similar
to the neutron spin rotation case, and consistent with a recent calculation\,\cite{Hyun:2006mp}.

The great interest of measuring $A_{\gamma}^{\vec{n}p}$ is mainly
because it is dominated by the OPE in the $V_{\textrm{OME}}^{\textrm{PV}}$
framework. This can also be observed from the above EFT analysis:
If one assumes the natural size of $\mN\rho_{t}/\wt{C}_{6}^{\pi}\sim0.1$,
the OPE contribution then dominates the SR one by a factor of $30$
or so.\,%
\footnote{We stress that this argument is based on naturalness. Without further
experimental confirmation, one should still keep other possibilities
open.%
} One of the outstanding puzzles in nuclear PV is the difficulty of
accommodating the extremely small upper limit on $h_{\pi}^{1}$, set
by the $^{18}\textrm{F}$ results~\cite{Adelberger:1985ik}, with
hadronic predictions and other nuclear PV experiments. The NPDGamma
experiment~\cite{Snow:2000az}, currently running at the Los Alamos
Neutron Science Center (LANSCE) and will be at SNS later on, aims
to reach an ultimate sensitivity of $5\times10^{-9}$. Results from
this experiment will certainly improve the long-existing value: $(0.6\pm2.1)\times10^{-7}$~\cite{Cavaignac:1977uk,Alberi:1988fd},
and hopefully resolve the $h_{\pi}^{1}$ puzzle.

Concluding this section, we shall make an important remark about the
PV meson exchange current (MEC) effects which manifest in electromagnetic
processes such as radiative neutron capture being discussed here and
in Sec.\, \ref{sub:P_g}. Although the Siegert theorem alleviates
much of the problem regarding the calculations of transverse electric
multipole operators $E_{J}$'s due to MECs, there is no easy simplification
when the transverse magnetic multipole operators $M_{J}$'s are concerned.
Furthermore, the Siegert theorem only applies to MECs which are constrained
by the continuity equation; for other transverse MECs, their effects
to $E_{J}$'s have to be added separately.

In Ref.\,\cite{Zhu:2004vw}, there is indeed such a transverse MEC,
which can not be accounted for by gauging the PV interaction, and
it introduces a new PV constant designated as $\bar{c}_{\pi}$. This
MEC takes the following form in the configuration space \begin{align}
\bm j_{\bar{c}_{\pi}}(\bm x;\bm x_{1},\bm x_{2})= & -i\,\frac{\sqrt{2}\, g_{\pi}\,\bar{c}_{\pi}}{\mN\,\Lambda_{\chi}\, F_{\pi}}\,(\tau_{1+}\,\tau_{2-})\left[\bm\sigma_{2}\cdot\hat{r}\,\bm\sigma_{1}\times\bm\nabla_{x}\,\delta^{(3)}(\bm x-\bm x_{1})\right]\nonumber \\
 & \times\frac{\mbox{e}^{-m_{\pi}\, r}}{4\,\pi\, r^{2}}\,(1+m_{\pi}\, r)+(1\leftrightarrow2)\,,\end{align}
where $\tau_{\pm}=\tau_{x}\pm i\,\tau_{y}$, and $r=|\bm x_{1}-\bm x_{2}|$.
Compared with the dominant part of the PV OPE MEC, the so-called pair
current, \begin{align}
\bm j_{\pi\mathrm{pair}}(\bm x;\bm x_{1},\bm x_{2})= & -\frac{g_{\pi}\, h_{\pi}^{1}}{2\,\sqrt{2}\,\mN}\,(\bm\tau_{1}\cdot\bm\tau_{2}-\tau_{1}^{z}\,\tau_{2}^{z})\left[\bm\sigma_{1}\,\delta^{(3)}(\bm x-\bm x_{1})\right]\nonumber \\
 & \times\frac{\mbox{e}^{-m_{\pi}\, r}}{4\,\pi\, r}+(1\leftrightarrow2)\,,\end{align}
the matrix element $\langle\bm j_{\bar{c}_{\pi}}\rangle$ roughly
scales with $\langle\bm j_{\pi\mathrm{pair}}\rangle$ by a factor
$\langle-i\,\bm\nabla_{x}/\Lambda_{\chi}\rangle=k/\Lambda_{\chi}$,
assuming $\langle r\rangle\sim1/m_{\pi}$ for typical nuclei. Therefore,
for the radiative processes considered in this work, where the photon
energy $k$ is just a few MeV so that $k/\Lambda_{\chi}\lesssim1\%$,
the contribution of $\bm j_{\bar{c}_{\pi}}$ is negligible. Hence
we do not have to include this extra PV constant $\bar{c}_{\pi}$
in the current search program at low energy.

\section{Summary~\label{sec:summary}}

In this work, we study the newly-proposed search program for nuclear
parity violation based on the effective field theory framework\,\cite{Zhu:2004vw}.
It is found that, in oder to completely describe the nuclear PV phenomena
at low energy, a minimal set of six parameters is needed. By low energy,
it means $E_{\mathrm{lab}}\lesssim90\,\mbox{MeV}$ for processes involving
$^{1}S_{0}$--$^{3}P_{0}$ transitions and $E_{\mathrm{lab}}\lesssim40\,\mbox{MeV}$
for ones involving $^{3}S_{1}$--$^{1}P_{1}$ and $^{3}S_{1}$--$^{3}P_{1}$
transitions. The six parameters to be determined phenomenologically
are the five dimensionless Danilov parameters: $\mN\lambda_{s}^{pp,nn,np}$,
$\mN\lambda_{t}$ and $\mN\rho_{t}$, and the long-range one-pion-exchange
parameter $\wt{C}_{6}^{\pi}$, which is proportional to the parity-violating
pion-nucleon coupling constant $h_{\pi}^{1}$. 

The two-body parity-violating observables being studied in this work
are summarized as following: \begin{align}
A_{L}^{\vec{p}p}(13.6\,\textrm{MeV}) & \approx-0.45\,\mN\lambda_{s}^{pp}\,,\\
A_{L}^{\vec{p}p}(45\,\textrm{MeV}) & \approx-0.78\,\mN\lambda_{s}^{pp}\,,\\
\frac{d}{d\, z}\,\phi_{n}^{\vec{n}p}(\textrm{th.})|_{\mathrm{rad/m}} & \approx0.30\,\wt{C}_{6}^{\pi}+2.50\,\mN\lambda_{s}^{np}-0.57\,\mN\lambda_{t}+1.41\,\mN\rho_{t}\,,\\
P_{\gamma}^{np}(\textrm{th.}) & \approx-0.16\,\mN\lambda_{s}^{np}+0.67\,\mN\lambda_{t}\approx A_{L}^{\vec{\gamma}d}(1.32\,\textrm{keV}+)\,,\\
A_{\gamma}^{\vec{n}p}(\textrm{th.}) & \approx-0.27\,\wt{C}_{6}^{\pi}-0.093\,\mN\rho_{t}\,.\end{align}
Because $A_{L}^{\vec{p}p}(13.6\,\textrm{MeV})$ and $A_{L}^{\vec{p}p}(45\,\textrm{MeV})$
essentially determine the same quantity, $\mN\lambda_{s}^{pp}$, these
equations only serve as four constraints---if precise data can all
be obtained for the three listed neutron experiments. In order to
have at least two more linearly-independent equations, other experimental
possibilities have to be explored. In few-body systems, where reliable
theoretical analyses can be performed, the candidate reactions include
$p\, d$, $n\, d$, $p\,\alpha$, $n\,\alpha$ etc.---just to name
a few. Currently, there are a published datum for $p\,\alpha$: $A_{L}^{\vec{p}\alpha}(45\,\mbox{MeV})=-(3.3\pm0.9)\times10^{-7}$\,\cite{Lang:1986nw},
and an ongoing experiment of thermal neutron spin ration in liquid
helium, $\phi_{n}^{\vec{n}\alpha}(th.)$, at the National Institute
of Standard and Technology\,\cite{Markov:2002}. In this respect,
existing calculations of these PV five-body processes should be updated.
Because alpha particle is a tightly bound state such that nucleons
inside have larger momenta, whether the $S$--$P$ dominance---the
cornerstone of this six-parameter analysis---can still hold should
be carefully examined. On the other hand, low-energy reactions involving
$d$ or $t$ might suffer less the problem. However, in order to motivate
new experiments, updated theoretical works are indispensable. 

\begin{acknowledgments}
The author would like to thank B.R. Holstein and M.J. Ramsey-Musolf
for the encouragement of taking on this project. The useful discussions
with them and J. Carlson, B. Desplanques, W.C. Haxton, and U. van
Kolck are deeply appreciated. Part of this work was supported by the
Dutch Stichting vor Fundamenteel Onderzoek der Materie (FOM) under
program 48 (TRI$\mu$P) and the EU RTD network under contract HPRI-2001-50034
(NIPNET). 
\end{acknowledgments}
\bibliographystyle{apsrev}
\bibliography{draft}

\end{document}